\documentclass[compsoc,conference,a4paper,10pt,times]{IEEEtran}
\IEEEoverridecommandlockouts
\usepackage{cite}
\usepackage{amsmath,amssymb,amsfonts}
\usepackage{algorithmic}
\usepackage{graphicx}
\usepackage{textcomp}
\usepackage{bmpsize}
\usepackage{xcolor}
\usepackage{lipsum}
\usepackage[colorlinks=true,urlcolor=black]{hyperref}
\usepackage{tikz}
\usepackage{multirow}
\usepackage{booktabs,xltabular} 
\usepackage{xspace}
\usepackage{tcolorbox}
\usepackage{algorithm}
\usepackage{mathtools}
\usepackage{array}
\usepackage{soul}
\usepackage{float}

\newcommand{\TheName}{\textsc{MalMixer}}

\def\BibTeX{{\rm B\kern-.05em{\sc i\kern-.025em b}\kern-.08em
    T\kern-.1667em\lower.7ex\hbox{E}\kern-.125emX}}
    
\newcommand{\squishlist}{
 \begin{list}{$\bullet$}
   { \setlength{\itemsep}{0pt}
     \setlength{\parsep}{0pt}
     \setlength{\topsep}{0pt}
     \setlength{\partopsep}{0pt}
     \setlength{\leftmargin}{2.5em}
     \setlength{\labelwidth}{1.5em}
     \setlength{\labelsep}{0.5em} } }

\newcommand{\squishend}{
  \end{list}  }
  
\makeatletter 
\newcommand{\linebreakand}{%
  \end{@IEEEauthorhalign}
  \hfill\mbox{}\par
  \mbox{}\hfill\begin{@IEEEauthorhalign}
}
\makeatother 

\sethlcolor{lightgray}

\begin{document}
%
\title{\TheName{}: Few-Shot Malware Classification with Retrieval-Augmented Semi-Supervised Learning}

\author{
\IEEEauthorblockN{
Jiliang Li\textsuperscript{†§*\thanks{*Equal contribution; authors listed in alphabetical order.}},
Yifan Zhang\textsuperscript{†*},
Yu Huang\textsuperscript{†},
Kevin Leach\textsuperscript{†}
}
\IEEEauthorblockA{\textsuperscript{†}Department of Computer Science, Vanderbilt University, Nashville, TN, USA\\}
\IEEEauthorblockA{\textsuperscript{§}Department of Computer Science, Stanford University, Stanford, CA, USA\\}
}

\maketitle

\begin{abstract}
Recent growth and proliferation of malware have tested practitioners' ability to promptly classify new samples according to malware families. 
In contrast to labor-intensive reverse engineering efforts, machine learning approaches have demonstrated increased speed and accuracy. 
However, most existing deep-learning
malware family classifiers must be calibrated using a large number of samples that are painstakingly manually analyzed before training.
Furthermore, as novel malware samples arise that are beyond the scope of the training set, 
additional reverse engineering effort must be employed to update the training set. The sheer volume of new samples found in the wild creates substantial pressure on practitioners' ability to reverse engineer enough malware to adequately train modern classifiers.

In this paper, we present \TheName{}, a malware family classifier using semi-supervised learning that achieves high accuracy with sparse training data.
We present a domain-knowledge-aware data augmentation technique for malware feature representations, enhancing few-shot performance of semi-supervised malware family classification. 
We show that \TheName{} achieves state-of-the-art
performance in few-shot  malware family classification settings. 
Our research confirms the feasibility and effectiveness of lightweight, domain-knowledge-aware data augmentation methods for malware features and shows the capabilities of similar semi-supervised classifiers in addressing malware classification
issues. 
\end{abstract}


%

\section{Introduction}\label{sec:intro}
Malware  has emerged as a paramount concern threatening software security, networks, and users.
In 2021, approximately $15.45\%$ of computers worldwide faced at least one malware attack~\cite{kaspersky2021bulletin} and the number of new malware variants continue to grow at an alarming rate.
For instance, compared to 2019, Windows business malware threats in 2021 increased by $85\%$ and consumer threats increased by $47\%$~\cite{malwarebytes2022}.
One important approach to describe malware behavior and consequences is classification into distinct families (e.g., sfone, wacatac, etc.).
Thus, prompt and accurate classification of malware is critical for effective defense against potential breaches.

In recent years, commercial antivirus scanners and machine-learning-based malware family classifiers have demonstrated high accuracy on test datasets~\cite{abusitta2021malware, aslan2020comprehensive}.
However, these classifiers face a practical challenge: classifiers fine-tuned on previously-seen samples do not work well on new samples~\cite{chen2017semi, gandotra2014malware, rad2011evolution}, as new malware samples, variants, and families with novel behaviors are constantly developed in the cat-and-mouse game between defenders and attackers~\cite{abusitta2021malware, gibert2020rise, pendlebury2021machine,raff2020survey}.

In the face of new, adaptive malware samples or zero-day threats, conventional malware family classifiers must be updated to handle new samples. 
Unfortunately, incorporating new malware samples requires substantial reverse engineering effort, which is both costly and time-consuming~\cite{santos2011semi,6620049,idika2007survey,mohaisen2013unveiling}, taking engineers with years of experiences hours to weeks to fully understand a single malware sample~\cite{votipka2020observational,raff2020survey}.
Thus, malware analyzers and machine learning solutions degrade in performance over time as new malware samples appear in the wild, requiring incremental retraining~\cite{narayanan2017context, xu2019droidevolver}.
This again requires effort from sufficiently-trained experts to analyze new samples to provide a basis for such retraining.

Since manual malware analysis and labeling are costly both in time and resources~\cite{chen2017semi}, when new malware samples appear in the wild, only a small portion can be labeled quickly enough to provide timely identification. Such identification is essential, as an unidentified malware (\texttt{DroidDream} in 2011) could infect as many as $260,000$ users with $48$ hours~\cite{grace2012riskranker}, and
training classifiers with promptly-acquired labels for previously-unseen samples can increase accuracy by up to $20\%$~\cite{miller2016reviewer}. 

To effectively use a small, promptly-obtained labeled dataset, one key challenge is that a small dataset may not be sufficiently diverse or representative of the entire cohort of related malware samples~\cite{wong2022marvolo}.
Thus, we desire techniques that enable robust and accurate classification of novel samples based on a relatively small corpus of manually-labeled malware samples --- that is, a \emph{few-shot} approach to robust malware family classification.

Previously, several papers have explored few-shot semi-supervised malware classifiers.
For instance, CAD-FSL~\cite{kasarapu2022cad} presents a framework using code-aware GANs to synthesize variants of labeled malware binaries, enhancing a CNN classifier.
Similarly, Catak et al.~\cite{catak2021data} employ image augmentation techniques like noise injection for generating new malware data.
These works 
use
\textit{data augmentation} to grow the size of input samples and enhance a downstream classifier.
As data augmentation is well-established in computer vision~\cite{wong2022marvolo}, almost all existing work employing augmentation first converts malware binaries to grayscale images (byteplots) on which to employ image augmentation. 
That said, image-based techniques do not modify malware feature representations in a \emph{semantically-meaningful} manner --- informally, varying pixel data does not correspond to sensible malware properties.
Compared to image-based techniques, semantics-preserving data augmentation methods exist, but typically require semantics-aware transformations on malware binaries~\cite{wong2022marvolo}, which can be very time and resource-intensive.
Thus, there is a need for data augmentation techniques that more sensibly augment malware feature representations by leveraging domain knowledge on the data's semantics, while also remaining computationally efficient for real-world applicability.
In designing our approach, we perform semantics-aware data augmentation on malware's feature representations, making a deliberate trade-off between some semantic validity and operational efficiency to achieve a practical solution.

\begin{figure}[tb]
\centering
\includegraphics[width=0.42\textwidth]{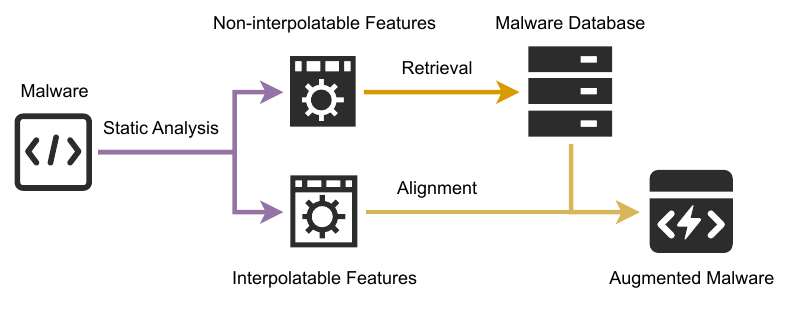}
\vspace{-8pt}
\caption{Overview of our proposed data augmentation approach for malware feature representations. We divide static malware features into \emph{interpolatable} and \emph{non-interpolatable}  sets, and use retrieval and alignment techniques to generate synthetic samples.}\label{fig:illustration}
\end{figure}

In this paper, we present \TheName{}, a few-shot malware classifier unifying 
(1) a retrieval-based, domain-knowledge-aware malware feature augmentation method that is aware of the underlying semantics of each malware feature, and 
(2) a semi-supervised learning framework that accentuates the effect of our augmentation approach.  
This combination provides robust and accurate family classification with few labeled malware samples.
At a high level, we grow a malware dataset through domain-knowledge-aware data augmentation on malware features. 
As shown in Figure~\ref{fig:illustration}, we divide malware features into \emph{interpolatable} and \emph{non-interpolatable} sets (defined in Section~\ref{sec:approach:interpolatable}) that are treated differently during training. 
Then, we use a similarity-based retrieval technique to mutate \emph{non-interpolatable} malware features and use manifold alignment to maximize the compatibility between the augmented \emph{interpolatable} and \emph{non-interpolatable}  features.
Finally, the augmented pseudo-malware samples are used to enhance a semi-supervised framework that combines consistency regularization, entropy minimization, and traditional regularization, following best practices in other domains~\cite{berthelot2019mixmatch}.
The resulting approach enables efficiently and accurately classifying large malware datasets by requiring only a few labeled samples of families of interest.

We leverage three key insights: 
First, malware samples within a given family share a majority of feature representations, despite variations on small subsets of features.
When labeled data is scarce, we can reasonably assume that samples most similar to the labeled data belong to the same malware family.
Second, two similar, same-family malware instances often differ only in semantics-preserving code transformations~\cite{wong2022marvolo}.
For example, two similar trojan variants may differ only by shallow static features (e.g., no-ops, junk code).  
As a result, while certain features like entropy and file size may differ, features that remain consistent between the two variants, such as opcode distribution in particular sections of the binary, may better reflect malicious behavior.
Given this insight, we perform augmentation by mixing existing feature representations of two similar, supposed same-family malware instances following interpolation rules, leading to improved capturing of essential invariant features (i.e., essential malicious behavior).
Third, many static analysis features of malware are interdependent.
For example, the size of the import table in a malware binary is closely related to the actual functions it imports.
This dependency between malware features indicate that, to generate synthetic malware features that meaningfully reflect plausible real-world samples, we must strive to modify each feature consistently 
with other related features. 
Thus, we use alignment techniques to ensure that mutated features in synthetic samples exhibit this interdependence.

We evaluate \TheName{} using BODMAS~\cite{yang2021bodmas} and MOTIF~\cite{joyce2022motif}, two popular benchmark datasets for malware family classification. 
Our evaluation focuses on 
\TheName{}'s efficacy in addressing scarce labeled data amidst influx of new malware instances, and is structured around four tasks: 
(1) saturation analysis to examine whether the classifier retains adequate performance with fewer labeled samples,  
(2) model comparison with other static-feature-based state-of-the-art models in few-shot settings, and evaluations of the \TheName{}'s readiness to handle new malware (3) originating from
different timeframes or (4) representing previously unseen families.
Our results show that \TheName{} surpasses current semi-supervised classifiers when dealing with limited labeled samples and temporal shift of malware data distributions (i.e., as existing families evolve). Specifically, in temporal analysis, \TheName{} surpasses current models by a relative margin of $11.44\%$.
The associated code and experimental data are publicly available via  \href{https://doi.org/10.5281/zenodo.8088356}{https://doi.org/10.5281/zenodo.8088356}. 

We claim the following contributions:  

\squishlist
    \item We introduce \TheName{}, a few-shot static-feature-based malware family classifier unifying domain-knowledge-aware data augmentation and semi-supervised learning. 
    \item We design a similarity-and-retrieval-based data augmentation technique on malware feature representations to generate synthetic data.
    \item We derive a manifold alignment technique to encourage consistency between all synthesized feature values to synthesize samples that mimic ground-truth family distributions.
    \item We thoroughly evaluate \TheName{}, showing we outperform other state-of-the-art static-feature-based malware classifiers in few-shot settings.
\squishend

\section{Background and Related Work}\label{sec:background}

In this section, we discuss (1) malware analysis methods to generate feature representations of malware, 
(2) data augmentation methods that grow training sets, (3) semi-supervised learning and (4) domain-invariant learning, from which we derive our classifier model.

\noindent\textbf{Malware Feature Analysis} 
Malware analysis can be divided into dynamic and static techniques~\cite{raff2020survey}, which in turn provide features that can be fed to a model.
Dynamic features are obtained by running malware in an instrumented environment to observe behavior~\cite{abusitta2021malware}.
In contrast, static features are extracted directly from 
malware binaries (e.g., Portable Executables (PEs) in Windows)~\cite{abusitta2021malware}. 
While dynamic analysis can reveal malicious behavior, it requires substantial time and computational resources~\cite{gandotra2014malware}.
In contrast, static features do not require execution of the sample, and can often be extracted quickly~\cite{raff2020survey}. 
In this paper, \TheName{} aims to enhance malware classification using static features extracted from malware PE binaries.
We use data augmentation to improve classification robustness against malware variants.

\noindent\textbf{Malware Data Augmentation}  Data augmentation is a technique used to expand a dataset by applying input transformations that do not alter the underlying class semantics~\cite{berthelot2019mixmatch}.
Well-established within computer vision communities, conventional augmentation uses mutations such as image zooming, cropping, and noise injection~\cite{shorten2019survey} to create novel samples that can be used to further train an underlying model.
Many security researchers have explored converting malware binaries into grayscale images, and subsequently applying image-based augmentation techniques~\cite{burks2019data, marastoni2021data, tekerek2022novel, kasarapu2022cad, catak2021data}.
Nevertheless, these image-based techniques are rather noise-inducing rather than  semantically meaningful to malware features.
Recently, Wong et al.~\cite{wong2022marvolo} demonstrated the feasibility of semantically-meaningful data augmentation of malware binaries.
The researchers performed semantics-preserving code transformations on labeled binaries to generate new mutated malware samples, therefore boosting the performance of ML-driven classifiers.
However, this is still time-consuming and resource-intensive and can only be practically conducted on a small number of samples~\cite{wong2022marvolo}.
In this paper, we present a lightweight feature-based, domain-knowledge-aware data augmentation technique, guided by the semantics of the static malware features.

\noindent\textbf{Semi-Supervised Learning} Semi-supervised learning is widely used to address data scarcity issues by combining limited labeled data with a volume of unlabeled data during the training process.
Popular strategies for semi-supervised learning progressively update the unlabeled dataset with high-confidence labeled data using techniques like label mixing and bootstrapping~\cite{arazo2020pseudo,zhang2021flexmatch}, curriculum learning~\cite{cascante2021curriculum,gong2016multi}, and meta-learning~\cite{pham2021meta,ren2018meta}.
Concurrently, they also aim to introduce regularization terms to ensure model consistency~\cite{zhang2017mixup,berthelot2019mixmatch,duarte2019semi}.
In malware classification, several studies have employed semi-supervised learning techniques, such as malware embedding~\cite{duarte2019semi,santos2011semi}, transfer learning~\cite{gao2020malware}, and soft labeling~\cite{mahdavifar2020dynamic}.
\TheName{} builds upon former works by introducing a novel data-centric augmentation method to enhance semi-supervised learning.

\noindent\textbf{Domain Invariant Learning} Domain Invariant Learning is a subfield of machine learning focused on generalizing models across different but related domains.
Researchers aim to identify \emph{invariant features} in data that characterize common trends amongst different domains that assist downstream model performance. One important category of Domain Invariant Learning is feature invariance learning, in which domain shift is reduced using unsupervised techniques to extract domain-invariant features~\cite{lu2022domain,xu2021anomaly}. 
In this paper, we use feature invariance learning to identify common malware features across malware families to better generate synthetic samples as part of a malware data augmentation pipeline. 

\section{Approach}\label{sec:approach}

We present \TheName{}, a few-shot malware family classifier unifying retrieval-based data augmentation and semi-supervised learning. 
We also introduce the underlying assumptions we make in constructing this approach.

Recall the goal of our approach is to identify the malware family of a given input malware sample. 
To achieve this, we train a classifier on a large dataset with only a small subset of expert-labeled malware samples. 
For each sample in the dataset, we apply static analyses to extract features. 
Then, within the feature space, we apply our data augmentation technique to each original sample in the dataset to generate more novel samples that closely reflect the distributions of the original samples.
Finally, we feed all augmented pseudo-malware samples into a semi-supervised classification framework that accentuates the effect of our proposed data augmentation.

Before proceeding to model training, each malware executable is 
statically analyzed (Section~\ref{sec:training}) to extract features. 
As such, \TheName{} is primarily geared towards enhancing static-feature-based malware classification. This domain is important for several reasons: 
(1) extracted features typically yield superior classification results compared to end-to-end classifiers that process entire malware binaries~\cite{anderson2018ember}, 
(2) static features can be promptly derived from new datasets, and
(3) the rise of sophisticated sandbox-evasion techniques has led to a shift to static analysis in industrial settings~\cite{aghakhani2020malware}.
Typical static analysis features include 
file size, hash of imported libraries, target OS, section information, byte entropy, among others~\cite{yang2021bodmas}. 
In general,  static features are not engineered for detecting certain families but rather provide a comprehensive summary of a PE file.
We evaluate \TheName{} on a set of state-of-the-art, statically extracted PE feature representations in the EMBER format~\cite{anderson2018ember}.

\begin{figure}[tbp]
\centering
\includegraphics[width=0.42\textwidth]{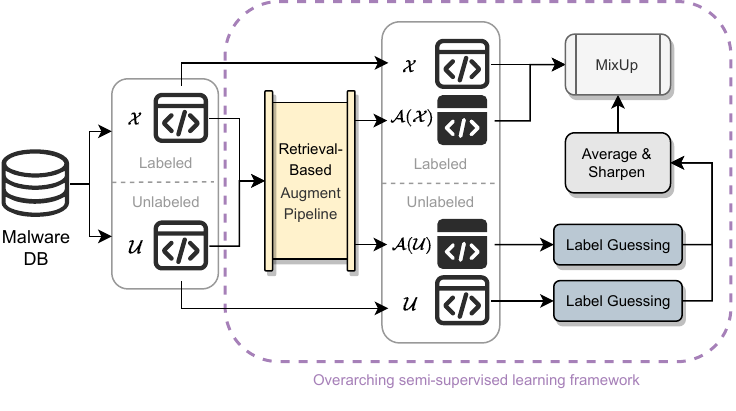}
\vspace{-8pt}
\caption{Illustration of our model framework, which consists of (1) a retrieval-based augmentation pipeline that grows the number of training samples and (2) an overarching augmentation-based semi-supervised classification framework that bolsters the augmentation pipeline. }
\label{fig:framework}
\end{figure}

Figure~\ref{fig:framework} illustrates \TheName{}, which consists of (1) a retrieval-based data augmentation pipeline, and 
(2) an overarching semi-supervised classification framework that accentuates the effect of the data augmentation pipeline.

\subsection{Semi-Supervised Framework} \label{sec:framework}

First, we present \TheName{}'s backbone: an data-augmentation-based semi-supervised classification framework. 
We use semi-supervised learning to build a malware classifier that requires only a small portion of the dataset to be labeled. 
In practice, our approach has the potential to reduce the expensive manual effort in reverse engineering and labeling malware samples. 

Our framework is based on MixMatch~\cite{berthelot2019mixmatch}, a semi-supervised learning framework originally designed for image classification. 
MixMatch transfers knowledge preserved in minimal labeling to largely unexplored domains.
Thus, we adapt MixMatch for malware family classification, allowing us to expand a small corpus of labeled malware to a much larger, unlabeled dataset.
MixMatch excels with semantics-preserving data augmentation, but traditional image-based augmentation techniques (e.g., adding noise to byteplots~\cite{malimg}) do not respect malware samples' semantics (Section~\ref{sec:background}). Thus, \TheName{} innovates upon prior art 
via domain-knowledge-aware augmentation tailored for malware data augmentation by considering malware features' underlying semantics. The semi-supervised framework and our augmentation approach can achieve outstanding classification accuracy and resiliency against unseen malware samples, as shown in Section~\ref{sec:analysis}.

We now introduce the overarching semi-supervised framework adopted from MixMatch. At a high level, unlabeled malware is combined with ground truth labeled malware through data augmentation, label guessing, averaging, sharpening, and MixUp~\cite{zhang2017mixup}, before training a base classifier. In this paper, we use a Fully Connected ResNet (FC-ResNet) as our base classifier (Appendix~\ref{sec:appendix:implementation}).

\noindent\textbf{Data Augmentation} 
We denote the set of labeled data as $\mathcal{X}$, their labels as $\mathcal{Y}$, and unlabeled data as $\mathcal{U}$. Note that $\mathcal{X}$ and $\mathcal{U}$ are feature representations of malware executables. First, we apply data augmentation to $\mathcal{X}$ and  $\mathcal{U}$ and transform them into $\mathcal{A}(\mathcal{X})$ and $\mathcal{A}(\mathcal{U})$,
where $\mathcal{A}$ is any stochastic data augmentation process s.t. 
\begin{equation}\label{eq:label}
    label(\mathcal{A}(s)) = label(s) ~\forall \text{~malware~} s\in \mathcal{X}\cup\mathcal{U},
\end{equation}
with $label$ being a function that provides the ground truth label for a malware sample. Our  augmentation pipeline will be introduced in Subsection~\ref{sub:information_retrieval}.

\noindent\textbf{Label Guessing, Average, Sharpening, and MixUp} 
Next, we generate pseudo labels for the unlabeled dataset. The base classifier $P$ predicts label probability distributions over all labels for both $\mathcal{U}$ and $\mathcal{A}(\mathcal{U})$. We take the average of these predictions to obtain more stable pseudo-label probabilities $\mathcal{Y_\mathcal{U}} = \frac{P(Y|\mathcal{U})+P(Y|\mathcal{A}(\mathcal{U}))}{2}$. Intuitively, this step encourages the classifier to assign consistent, similar labels to the original and augmented versions of the same malware.

We then sharpen pseudo-label  probabilities $\mathcal{Y_\mathcal{U}}$ with $\text{Sharpen}(y)_i = \frac{y_i^2}{\sum_j y_j^2}$. This step is akin to applying softmax on pseudo-label distributions to reduce uncertainty by enhancing confident label predictions. 

These steps provide us pseudo-label probabilities for each original unlabeled sample and its corresponding augmented synthetic sample.
Combining the labeled dataset and the pseudo-labeled unlabeled dataset, we construct a fully labeled dataset $\mathcal{X^{\prime}} = (\mathcal{X}\cup\mathcal{A}(\mathcal{X}), \mathcal{Y}\cup\mathcal{Y})$ and $\mathcal{U^{\prime}} = (\mathcal{U}\cup\mathcal{A}(\mathcal{U}),\mathcal{Y_\mathcal{U}}\cup\mathcal{Y_\mathcal{U}})$. 

 Next, we apply MixUp~\cite{zhang2017mixup} to labeled and pseudo-labeled samples. MixUp performs linear interpolation of features and label probabilities, giving us
$\mathcal{X^{\prime}} = M(\mathcal{X^{\prime}}, \mathcal{W_{X^{\prime}}}), ~\mathcal{U^{\prime}} = M(\mathcal{U^{\prime}}, \mathcal{W_{U^{\prime}}})$. Here, $M$ is linear interpolation and $\mathcal{W_{X^{\prime}}}, \mathcal{W_{U^{\prime}}}$ are randomly selected subsets of the dataset, so that every sample is mixed up with some other sample. Intuitively, if feature representation $[0,0]$ has label probabilities $[1,0,0]$ and and feature $[2,2]$ has label $[0,1,0]$, MixUp encourages the classifier to assign label probabilities $[\frac{1}{2},\frac{1}{2},0] = M([1,0,0], [0,1,0]) = \frac{1}{2}[1,0,0]+ \frac{1}{2}[0,1,0]$ to feature $[1,1] = M([0,0], [2,2]) = \frac{1}{2}[0,0]+ \frac{1}{2}[2,2])$. This encourages the classifier to behave linearly between samples, smoothing decision boundaries and improving robustness. 

To train a base classifier, the loss is  
\begin{align*}
    \mathcal{L_{X^\prime}} &= \frac{1}{|\mathcal{X^{\prime}}|} \sum_{x, y \in \mathcal{X^{\prime}}}{\text{CrossEntropy}(y, P(y|x))},\\
    \mathcal{L_{U^\prime}} &= \frac{1}{|\mathcal{U^{\prime}}|} \sum_{u, y \in \mathcal{U^{\prime}}}{||y - P(y|u)||_2^2},
\end{align*}
The final loss is $\mathcal{L} = \mathcal{L_{X^\prime}} + \mathcal{L_{U^\prime}}$.
The trained base classifier $P$ is then used to classify test sets.

\begin{figure*}[t]
\centering
\includegraphics[width=0.92\textwidth]{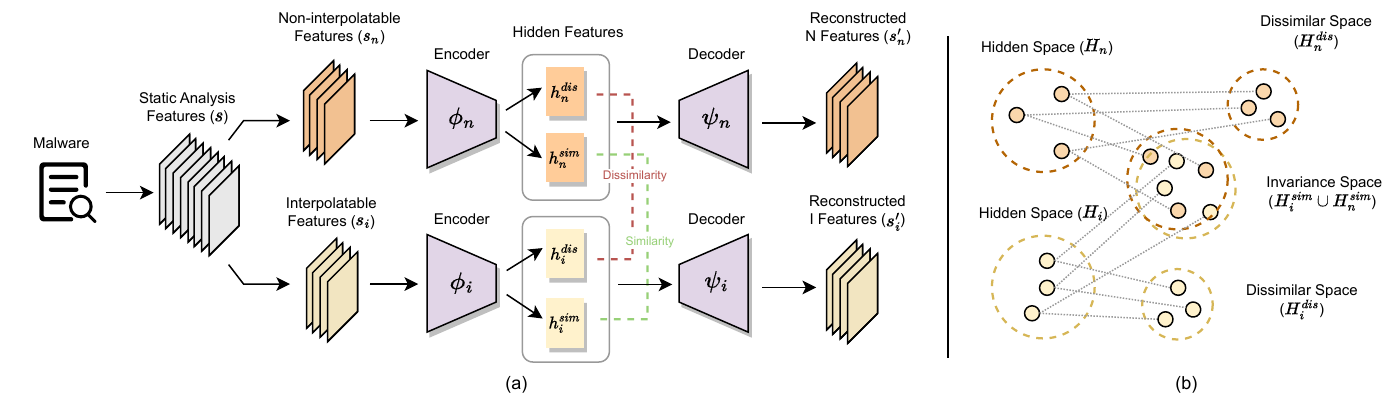}
\vspace{-8pt}
\caption{Diagram illustrating Domain Invariance Learning for malware. As shown in part (a), the process contains three components: malware reconstruction, similarity learning, and dissimilarity learning. The static analysis features of malware are categorized into interpolatable ($s_i$) and non-interpolatable ($s_n$) features. During malware reconstruction, an encoder-decoder architecture accepts $s_i$ and $s_n$ as inputs and reconstructs them into $s_i^\prime$ and $s_n^\prime$. Simultaneously, the architecture's hidden features are divided into two sections, each forced to learn invariant and dissimilar features, respectively. Part (b) visualizes the learned hidden features, projecting malware into dissimilar spaces and an intersecting invariance space. 
\vspace{-8pt}
}\label{fig:encoder}
\end{figure*} 
\begin{figure}[tbp]
\centering
\includegraphics[width=0.44\textwidth]{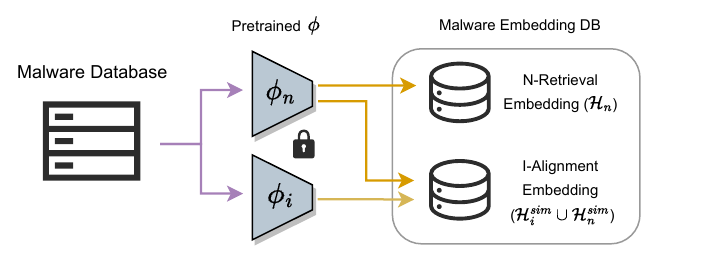}

\caption{Illustration of embedding projection for malware. Pretrained encoder models transform each malware feature representation into embeddings for N-Retrieval and I-Alignment purposes. 
}
\label{fig:retrieval}
\end{figure}

\subsection{Retrieval-Based Data Augmentation Pipeline}\label{sub:information_retrieval}
Within the aforementioned semi-supervised framework, a key component is data augmentation, $\mathcal{A}$, used in Equation~(\ref{eq:label}). 
In malware family classification, it is challenging to completely satisfy Equation~(\ref{eq:label}) due to difficulties with analyzing and reasoning about raw binaries~\cite{wong2022marvolo}. 
In this subsection, we introduce a data augmentation pipeline that approximates the equality in Equation~(\ref{eq:label}) using high-confidence,  domain-knowledge-aware mutations of each malware instance, denoted $s$.
The outcome is a set of synthetic malware samples that are well-grounded and reflect more realistic, plausible samples. 
The augmentation process is decoupled from our semi-supervised framework and can be performed prior to training.

Our proposed data augmentation pipeline mutates a malware sample by interpolating 
its features with a similar sample likely to be of the same family. 
However, the key challenge is straightforward interpolations are not necessarily semantically meaningful --- for example, taking the average of two MD5 hash values of strings in malware samples will not provide a meaningful, novel value. 

To address the challenge, we follow four steps:  
(i) separating malware features into \emph{interpolatable} and \emph{non-interpolatable} sets, 
(ii) employing domain-invariant learning to project static features of malware into embedding spaces, 
(iii) applying a retrieval-based method to identify candidate features representative of a mixture of two malware instances, 
(iv) using an alignment-based approach to generate the best synthetic malware sample whose interdependent features are most compatible.
Steps (i) and (ii) generate the oracles necessary for steps (iii) and (iv) that carry out the actual data augmentation.

\noindent\textbf{(i) Interpolatable vs. Non-Interpolatable Features}\label{sec:approach:interpolatable}

We first divide static features into \emph{interpolatable} (denoted $\mathcal{I}$) and \emph{non-interpolatable} (denoted $\mathcal{N}$).
The designation of interpolatable vs.  non-interpolatable follows an intuitive rule: a feature is interpolatable if and only if that feature is numerical in nature and perturbing that feature still results in sensible values.
For example, the byte entropy of a file is interpolatable since any interpolated value remains a plausible entropy for some executable.
Conversely, a hash of an imported filename is non-interpolatable, as not all hashes correspond to valid filenames.
As such, typical interpolatable features include byte entropy histograms, Shannon's entropy of printable characters, and the occurrence count of string literals.
Typical non-interpolatable features encompass major OS versions and hash values of libraries / section names.
The rule for splitting features into interpolatable vs non-interpolatable does not change based on features: for example, if using control-flow-graph features, those features non-numerical in nature (e.g., control dependencies) will be non-interpolatable.


With domain knowledge and research effort, the designation of interpolatable vs non-interpolatable features can be performed quickly on any arbitrary malware feature representations: for instance, in our studies, we processed a 2381-dimensional static feature representation of malware within five hours by an author with a security background. The set of features used in this study is the EMBER raw features~\cite{anderson2018ember}, which are used in the BODMAS and MOTIF datasets introduced in Section~\ref{sec:training}. The particular interpolatable vs. non-interpolatable split that 
we used for our experiments is described in Appendix~\ref{sec:appendix:interpolatable}.

We note that, when interpolating all features in the same way (i.e., without splitting them), classification performance degraded (Section~\ref{sec:appendix:ablation}).

As illustrated in Figure~\ref{fig:encoder} (a), 
the static features of each malware sample $s$ are divided into interpolatable ($s_i$) and non-interpolatable ($s_n$) features, such that $s=s_i \oplus s_n$.
For notation, we let $\mathcal{S} := \mathcal{X}\cup\mathcal{U}$ be the set of all malware $s$.
Alternatively, we can express $\mathcal{S} = concat(\mathcal{S}_i,\mathcal{S}_n)$, where $\mathcal{S}_i$ is the set of all interpolatable features $s_i$, $\mathcal{S}_n$ is the set of all non-interpolatable features $s_n$. $concat$ denotes the pairwise feature concatenation of $(s_i, s_n)$ derived from the same sample.

Parts (ii)–(iv) below present the mathematical formulations of the next steps in our proposed pipeline. However, readers may find it helpful to first refer to motivations behind our design in Section~\ref{sec:results-rq0} and/or a case study in Section~\ref{sec:case} that intuitively walks through steps (i)–(iv).

\begin{figure*}[tbp]
\centering
\includegraphics[width=0.88\textwidth]{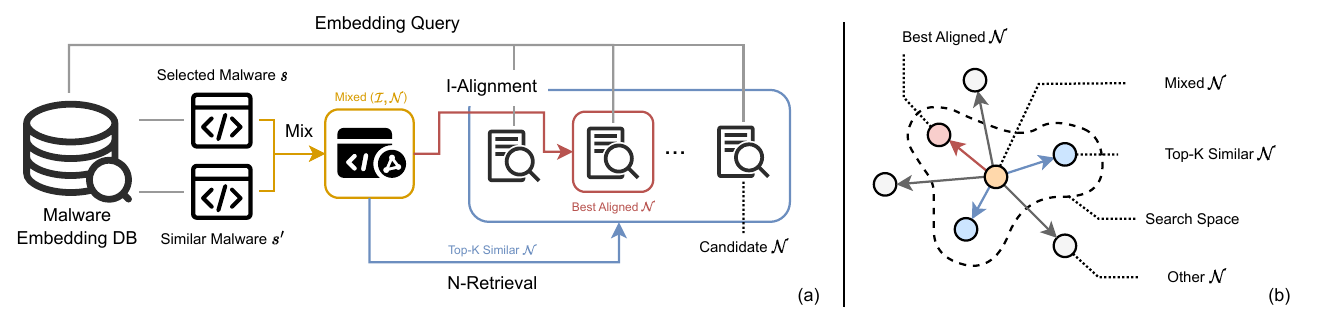}
\vspace{-8pt}
\caption{
Illustration of the retrieval and alignment process for identifying the optimal non-interpolatable $\mathcal{N}$ features for synthetic malware.
The diagram contains three components in part (a): malware mixing, N-retrieval, and I-alignment.
In malware mixing, two similar malware samples are combined based on their $\mathcal{N}$ and $\mathcal{I}$ features.
The mixed $\mathcal{N}$ features are used to search for the top-k similar $\mathcal{N}$ in the database to represent the mixed $\mathcal{N}$.
The mixed $\mathcal{I}$ features are then used to align with candidate $\mathcal{N}$ features and select the best aligned $\mathcal{N}$.
Part (b) visualizes the process of choosing a set of $\mathcal{N}$ features with the highest degree of alignment from all candidates in the invariance embedding.
}\label{fig:search} \end{figure*} 

\noindent\textbf{(ii) Invariance and Dissimilar Learning}

Next, we project malware features into two embeddings: (1) an invariant embedding, capturing the commonalities between interpolatable and non-interpolatable features from the same malware sample;
and (2) a dissimilar embedding, emphasizing the unique aspects that differentiate interpolatable from non-interpolatable features.
We use these embeddings to capture interdependent features and family-common features that are used to guide the data augmentation in steps (iii) and (iv).

To achieve this projection into two embedding spaces, we design a separate encoder-decoder structure for each of $s_i$ and $s_n$. Then, we perform feature reconstruction using
${s_i}^\prime = \psi_i(\phi_i(s_i)), ~{s_n}^\prime = \psi_n(\phi_n(s_n)),$
where $\phi_i$ (and $\phi_n$) and $\psi_i$ (and $\psi_n$) represent the encoders and decoders for $s_i$ (and $s_n$).
The reconstructed variables are denoted as ${s_i}^\prime$ and ${s_n}^\prime$. The corresponding loss for malware feature reconstruction is
$\mathcal{L_R} = \frac{1}{N} \sum_{k=1}^{N} \left( ||s_i - {s_i}^\prime||^2_2 + ||s_n - {s_n}^\prime||^2_2 \right)_k,$
where $\mathcal{L}_\mathcal{R}$ represents the reconstruction loss, and $N$ denotes the total number of data points in a batch. 
The encoder-decoder architecture enables transforming static features of malware into a more compact representation capturing the underlying structure and semantics of the executable.

To separate domain-specific features from domain-invariant features during the reconstruction process, we introduce two auxiliary tasks for the encoder-decoder. 
As shown in Figure~\ref{fig:encoder}, we divide the encoded hidden features $h_i\coloneqq\phi_i(x_i)$ and $h_n\coloneqq\phi_n(x_n)$ evenly, splitting them into two parts each. 
For instance, if the original dimension of $h_i$ is 512, each resulting vector will have a dimension of 256. We then perform similarity and dissimilarity learning on these vectors. 
Specifically, we define the two parts of the split vector as the \emph{similar vectors}, denoted as $h_i^{sim}$ and $h_n^{sim}$, and the \emph{dissimilar vectors}, denoted as $h_i^{dis}$ and $h_n^{dis}$, obtaining $h_i = h_i^{sim} \oplus h_i^{dis}$ and $h_n = h_n^{sim} \oplus h_n^{dis}$. 
We enforce the dissimilar vectors to be distant from one another, while ensuring the similar vectors remain close together. 
This is expressed as $\mathcal{L_S} = \frac{1}{N} \sum_{k=1}^{N} || h_i^{sim} - h_n^{sim} ||_2^2, ~\mathcal{L_D} = \frac{1}{N} \sum_{k=1}^{N} \left( \max \left(0, 5 - || h_i^{dis} - h_n^{dis} ||_2^2 \right) \right)_k,$
where $\mathcal{L_S}$ and $\mathcal{L_D}$ denote similarity loss and dissimilarity loss, respectively. The total loss is $\mathcal{L_R} + \mathcal{L_D} + \mathcal{L_S}$.
    

After training the encoder-decoder model, we lock the parameters of both encoders $\phi_i$ and $\phi_n$. 
Then, we apply $\phi_i$ to $\mathcal{S}_i$, $\phi_n$ to $\mathcal{S}_n$, and obtain an embedding of interpolatable features $\mathcal{H}_i:=\phi_i(\mathcal{S}_i)$ and an embedding of non-interpolatable features $\mathcal{H}_n:=\phi_n(\mathcal{S}_n)$. 
Note that $\mathcal{H}_i=concat(\mathcal{H}_i^{sim},\mathcal{H}_i^{dis})$, where $\mathcal{H}_i^{sim}$ is the invariant embedding of interpolatable features containing all $h_i^{sim}$, and $\mathcal{H}_i^{dis}$ is the dissimilar embedding of interpolatable features containing all $h_i^{dis}$. 
$concat$ is the pairwise feature concatenation of $(h_i^{sim}, h_i^{dis})$ such that $h_i^{sim}, h_i^{dis}$ belong to the same $h_i$. 
Similarly, $\mathcal{H}_n=concat(\mathcal{H}_n^{sim},\mathcal{H}_n^{dis})$.

As shown in Figure~\ref{fig:retrieval}, we refer to $\mathcal{H}_n$ as the N-retrieval embedding, and $\mathcal{H}_i^{sim}\cup\mathcal{H}_n^{sim}$, which is the invariant embedding of both interpolatable and non-interpolatable features, as I-alignment embedding. 
Note that a one-to-one correspondence exists between $\mathcal{S}$ (the original malware feature representation set), the N-retrieval embedding (representing non-interpolatable features), and the I-alignment embedding (representing common aspects between interpolatable and non-interpolatable features).
For every original malware sample in $\mathcal{S}$, there is a projection of its non-interpolatable features in the N-retrieval embedding.
Similarly, the shared aspects between each sample's interpolatable and non-interpolatable features are mapped onto the I-alignment embedding.

Once the N-retrieval and I-alignment databases are obtained, we proceed to steps (iii) and (iv), which mutate a malware instance by mixing it with a similar instance using these databases. The motivations for establishing these databases, as well as using them in steps below, are presented in our insights in Section~\ref{sec:intro} and Section~\ref{sec:results-rq0}. 
Steps (iii) and (iv) are illustrated by Figure~\ref{fig:search}.

\begin{figure*}[tbp]
\centering
\includegraphics[width=0.92\textwidth]{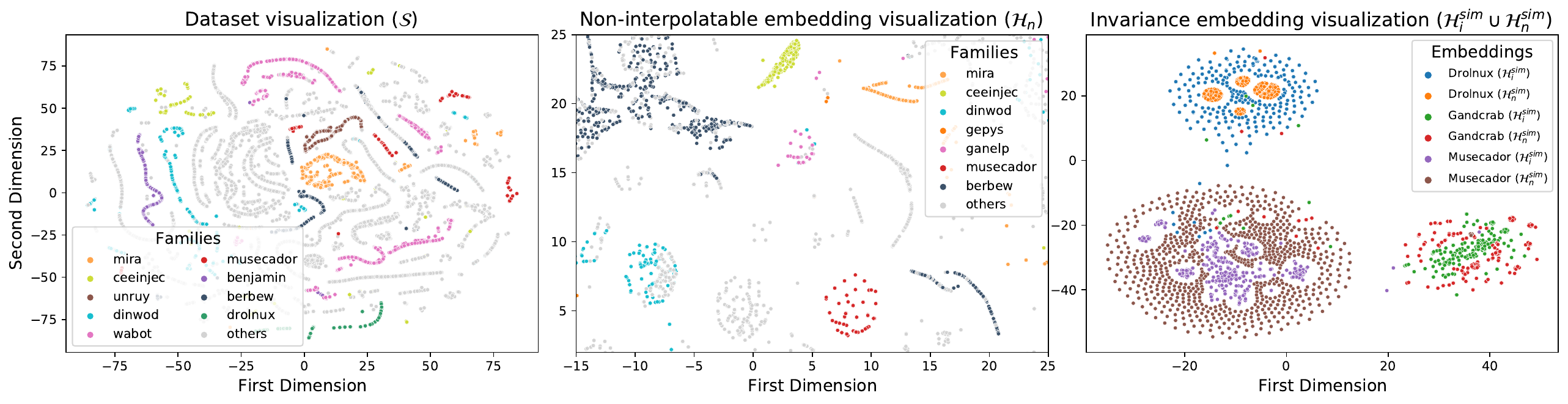}
\vspace{-8pt}
\caption{Illustration of similarities shared among families in BODMAS. 
The left subfigure displays such similarities in the original static feature space $\mathcal{S}$. The middle subfigure depicts shared similarities in the N-Retrieval embedding $\mathcal{H}_n$. 
The right subfigure represents similarities between interpolatable and non-interpolatable features of similar malware when projected to the I-Alignment invariance embedding $\mathcal{H}_i^{sim}\cup\mathcal{H}_n^{sim}$.}\label{fig:embedding_vis}\label{figure:similarity}
\end{figure*}

\noindent\textbf{(iii) Retrieval-Based Malware Matching}\label{sub:matching_and_retrieval}

To augment a malware instance, denoted $s$, step (iii) starts by finding several candidate features that are representative of mixing $s$ with another similar instance. 
To achieve this, three tasks must be performed: we 
(a) identify a similar malware sample $s^\prime$ in $\mathcal{S}$;
(b) linearly interpolate $s$ and $s^\prime$ by their interpolatable features; and
(c) select candidates representative of a mixture of $s$ and $s^\prime$ by their non-interpolatable features. 

For task (a), we identify the top-k similar malware samples in $\mathcal{S}$ compared to $s$ through constructing a k-nearest-neighbor graph. We select L2 distance computed on initial features as the similarity metric for the $k$-NN graph. 
From the top-k similar malware samples, we randomly select one neighbor, denoted $s^\prime$. 

Then, for task (b), we combine $s_i$ and $s^\prime_i$ (the interpolatable features of $s$ and $s^{\prime}$) to obtain 
\begin{equation}\label{eq:mix-I}
    \Tilde{s}_i := \alpha s_i + (1-\alpha)s_i^\prime.
\end{equation}

\noindent where $\alpha$ is a random number from 0 to 1. 
Note that by choice of linear combination, we only mutate features for which $s$ and $s^\prime$ are not shared. 

Finally, for task (c), we retrieve the $h_n, h_n^\prime$ corresponding to $s, s^\prime$ from $\mathcal{H}_n$. 
We mix $s$ and $s^\prime$ by their non-interpolatable features in the N-Retrieval embedding as
\begin{equation}\label{eq:mix-U}
    \Tilde{h}_n := \alpha h_n + (1-\alpha)h_n^\prime.
\end{equation}
where $\alpha$ is the same $\alpha$ used above. 
We consequently select the top-k nearest neighbors of $\Tilde{h}_n$ in $\mathcal{H}_n$ using L2 distance as similarity metric.
The selected neighbors, denoted $\{\hat{h}_{n_1},\cdots, \hat{h}_{n_k}\}$, are regarded as candidates in the N-Retrieval embedding space representing a mixture of non-interpolatable features of $s$ and $s^\prime$. 
A more detailed motivation for performing task (c) and the following step (iv) in the embedding spaces, rather than of the original space, will be provided in Section~\ref{sec:results-rq0}.

\noindent\textbf{(iv) Alignment-based malware selection} 

As we obtain candidates representative of mixing $s$ and $s^\prime$ by their non-interpolatable features, we then select among these candidates a set of non-interpolatable features that is most compatible with the mixed interpolatable features $\Tilde{s}_i$. 

First, we apply the pre-trained encoder $\phi_i$ to $\Tilde{s}_i$ and obtain projection $\Tilde{h}_i=\phi_i(\Tilde{s}_i)$. Recall that projection $\Tilde{h}_i$ can be broken down into similar and dissimilar vectors, i.e. $\Tilde{h}_i = \Tilde{h}_i^{sim}\oplus \Tilde{h}_i^{dis}$. Similarly, for each $\hat{h}_{n_l}\in \{\hat{h}_{n_1},\cdots, \hat{h}_{n_k}\}$, $\hat{h}_{n_l} = \hat{h}_{n_l}^{sim}\oplus \hat{h}_{n_l}^{dis}$. Thus, we perform alignment by selecting $\hat{h}_{n_j}$ among candidates $\{\hat{h}_{n_1},\cdots, \hat{h}_{n_k}\}$ such that 
$$j = argmin_l ~||\Tilde{h}_i^{sim} - \hat{h}_{n_l}^{sim}||, ~\hat{h}_{n_l}^{sim}\in \{\hat{h}_{n_1}^{sim},\cdots, \hat{h}_{n_k}^{sim}\}.$$ 
In doing so, we choose one set of non-interpolatable features that aligns most closely with the mixed interpolatable features in the I-alignment embedding. Recall that $\hat{h}_{n_j}$ has a one-on-one correspondence with some $\Tilde{s}_n\in\mathcal{S}_n$ in the original feature space. As a result, we use $\Tilde{s}_n$ as the mixture of $s$ and $s^\prime$ by their non-interpolatable features. 
We define our data augmentation method $\mathcal{A}$ on $s\in\mathcal{S}$ such that
 $\mathcal{A}(s) = \Tilde{s}_i\oplus \Tilde{s}_n$. 
 
In essence, $\mathcal{A}(s)$ is a concatenation of a direct mutation on the interpolatable features of $s$ and a real set of non-interpolatable features retrieved from another training sample. Steps (iii) and (iv) address the problem that arises when mixing non-interpolatable features from two samples, which typically results in a semantically nonsensical mixture. This issue is mitigated by selecting non-interpolatable features from real samples that are akin to the mixed instance (step (iii)) and compatible with the mutated interpolatable features (step (iv)). 

During the data augmentation process, establishing a one-on-one between space $\mathcal{S}$ and the hidden spaces $\mathcal{H}_i, \mathcal{H}_n$ allows the model to optimize across both the original interpolatable feature space and the non-interpolatable hidden space. This results in a more realistic synthetic sample. We provide a detailed justification for our design choices in the next subsection and a intuitive walkthrough of our proposed data augmentation pipeline in Section~\ref{sec:case}.
The data augmentation method is formulated in Algorithm~\ref{alg:algorithm} in the Appendix.

\begin{figure*}[tb]
\centering
\includegraphics[width=0.94\textwidth]{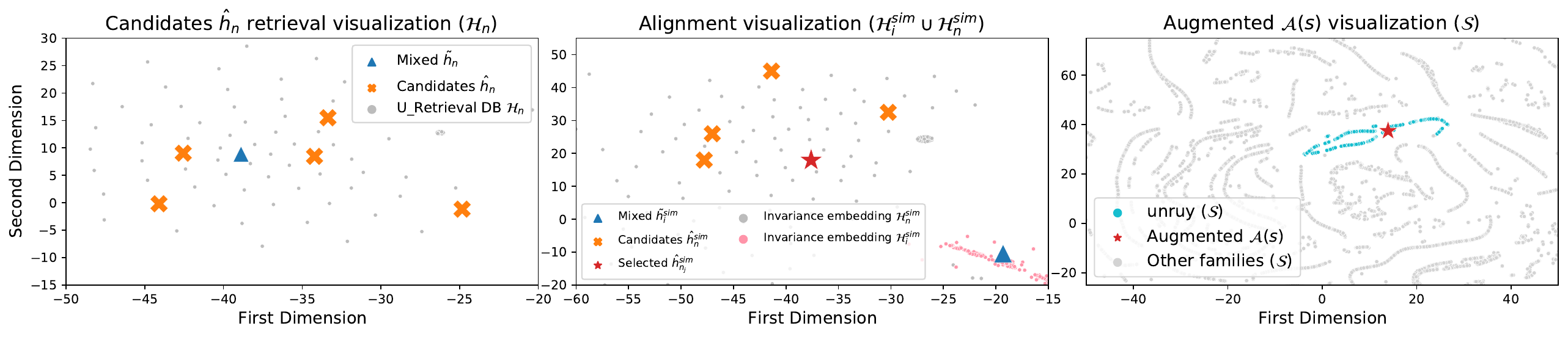}
\vspace{-8pt}
\caption{
Case study workflow.  In the left subfigure, we identify several non-interpolatable features representing a blend of the two chosen samples in the embedding space. In the middle subfigure, we select the non-interpolatable features that best align with the combined interpolatable features in the embedding. In the right subfigure, we map the synthetic malware, consisting of the chosen mixed features, back to the original feature space.
}\label{fig:case_visualization}
\end{figure*}

\subsection{Assumptions and Insights}\label{sec:results-rq0}
We make three key assumptions that underpin our data augmentation approach based on the insights described in Section~\ref{sec:intro}.  In this subsection, we describe and validate these assumptions.

(i) We assume that same-family malware are similar in the original static feature space ($\mathcal{S}$).
Consequently, we mutate each malware sample $s$ by mixing it with one of its top-k neighbors, presuming they are instances of the  same family and thus semantically similar (see Section~\ref{sub:information_retrieval} part (iii) task (a)).
We hypothesize that mixing two such samples likely yields a new sample with more realistic features than existing data augmentation techniques.

(ii) We assume that non-interpolatable features from similar malware are projected closer together within an embedding space.
Thus, we expect that mixing these features' projections in N-Retrieval embedding $\mathcal{H}_n$ via linear combination will neighbor other non-interpolatable features corresponding to similar functionality (see Section~\ref{sub:information_retrieval} part (iii) tasks (b) \& (c)). 

(iii) We assume that using alignment techniques, interpolatable and non-interpolatable features from the same malware are projected onto similar subspaces within an embedding space. 
Thus, similarity within the I-Alignment invariance embedding $\mathcal{H}_i^{sim}\cup\mathcal{H}_n^{sim}$ indicates compatibility between a pair of interpolatable and non-interpolatable features (i.e., some pairs of features may be interdependent).
Accordingly, we use similarity search between interpolatable and non-interpolatable features in this embedding space to align them and maximize feature consistency within samples we synthesize (see Section~\ref{sub:information_retrieval} part (iv)).

\noindent\textbf{Similarity analysis} 
Figure~\ref{figure:similarity} visualizes these insights using T-SNE~\cite{van2008visualizing} over samples in the BODMAS dataset.  
We measure similarity using Euclidean distance~\cite{van2008visualizing}. 
Assumption (i) is supported by the organized clusters of malware families in Figure~\ref{figure:similarity}-left.
Numerically, we also verified that, within BODMAS-20, a subset of BODMAS detailed in Section~\ref{sub:motivation-dataset-pre-processing}, each malware sample shares the same family as $95.86\%$ of its top-5 nearest neighbors. 
For assumption (ii), Figure~\ref{figure:similarity}-middle shows non-interpolatable features form family-based clusters in the N-Retrieval embedding, $\mathcal{H}_n$.
In contrast, most non-interpolatable features do not form clusters in the original feature space $\mathcal{S}_n$.
Thus, we mix non-interpolatable features in the embedding space $\mathcal{H}_n$ (Equation~(\ref{eq:mix-U})).
Finally, assumption (iii) is supported by the overlapping subspaces of same-family features in the invariance embedding $\mathcal{H}_i^{sim}\cup\mathcal{H}_n^{sim}$, as illustrated in Figure~\ref{figure:similarity}-right.
Consequently, similarity between projected interpolatable and non-interpolatable features indicates higher compatibility.

Note that our approach operates primarily in the projected embedding space. Our assumptions and similarity analysis demonstrate our manipulations yield meaningful relationships in this space, but they do not fully translate to an explanation of specific malware features, traits, or functionalities. In this regard, this Section as well as Section~\ref{sec:case} below, provide preliminary evidence that our approach captures relevant semantic patterns, rather than as an attempt to achieve full explainability of the underlying problem domain.

\section{Motivating Example}\label{sec:case}
The three assumptions described in Section~\ref{sec:results-rq0} underpin the logic behind several key steps in our data augmentation pipeline. 
Next, we provide a case study in which each step of our data augmentation pipeline is discussed.
In doing so, we aim to intuitively explain how each step works together to form novel, domain-knowledge-aware synthetic malware samples. Such augmented synthetic samples can then effectively enhance the performance of the overarching semi-supervised framework.

We first consider a pair of similar real malware Samples A and B, where sample A is labeled, but B is not.
Both samples belong to the same family, \texttt{unruy}, a family of trojan adware samples. 
Conventional data augmentation approaches that mix or perturb samples only consider feature values without regard for the underlying semantics of those features.
For example, each of Sample A and B contains a MurmurHash3 hash of the loaded libraries.
Naive approaches mutate these samples by averaging or adding noise to these hash values, resulting in values that do not correspond to sensible input strings.
  
Thus, to address this challenge, we first mix only the interpolatable features of A and B, such as entropy and string counts.
For non-interpolatable features like hash values, we retrieve real values from existing samples that (1) represent a mix of values from A and B in the embedding space, and (2) are consistent with the mixed interpolatable values obtained earlier.
The processes used to retrieve such values are visualized in the left and middle subfigures of Figure~\ref{fig:case_visualization}.
These processes ensure the newly-generated sample contains realistic features, such as library hash values and operating system versions, which can be traced back to known-real values, and that correspond to the computed interpolatable feature values.

Note that Figure~\ref{fig:case_visualization}-right shows how the newly-generated sample maps back to the \texttt{unruy} family cluster in feature space ($\mathcal{S}$). This shows that our synthetic sample accurately mimics malware in the \texttt{unruy} family.

We further analyze the synthetic sample generated by combining A and B.
The retrieved hash values of imported libraries trace back to a set of libraries that include KERNEL32.dll and USER32.dll, which are common Windows libraries.
The retrieved values for linker version and operating systems indicate the file is designed to run on 32-bit Windows systems, which is the primary target of the \texttt{unruy} family. 

We also assess the compatibility between interpolatable and non-interpolatable features of the synthetic sample, providing intuition for our alignment method. A substantial number of key interpolatable features of the synthetic sample (e.g., the number of import functions, import table size, import address table size, and MEM\_WRITE, MEM\_READ, and MEM\_EXECUTE operations) match exactly those of the real sample from which we retrieved its non-interpolatable features. 
Other features, such as Shannon's entropy of printable characters and the byte histogram, differ by only around $10\%$ from the real sample. 
This suggests that our data augmentation method creates a synthetic sample that contains realistic values for a novel and plausible \texttt{unruy} executable.

In addition to this intuitive example, we also conduct an Ablation Study in Section~\ref{sec:appendix:ablation} to empirically showcase how each step in our proposed augmentation pipeline contributes to accurate few-shot malware classification.

\section{Experimental Settings}\label{sec:experiment}
In this section, we 
present an overview of the datasets we use to evaluate \TheName{}. 
We also include our model architectures, implementation, training hyperparameters, and hardware resources in Appendix~\ref{sec:appendix:implementation}.

\subsection{Training Datasets}
\label{sec:training}

We train and evaluate \TheName{} on two datasets: (1) Blue Hexagon Open Dataset for Malware Analysis (BODMAS)~\cite{yang2021bodmas} and (2) the Malware Open-source Threat Intelligence Family (MOTIF) dataset~\cite{joyce2022motif}. 
These datasets are popular, up-to-date, and well-curated dataset baseline for PE malware familial classification tasks in recent research~\cite{someya2023fcgat, wu2023grim, dambra2023decoding, corlatescu2024embersim}. 

The BODMAS dataset contains $57,293$ malware samples representing $581$ families, obtained and timestamped between August 2019 and September 2020. The MOTIF dataset contains $3,095$ disarmed PE malware samples from $454$ families, published between January 2016 and January 2021.
All samples in both datasets are represented by a 2381-dimensional feature representation statically extracted from Windows PE binaries. The static features encompass byte entropy histogram~\cite{saxe2015deep}, string entropy, PE header information, sections table information, and hashed imported libraries, among others. The set of features in BODMAS and MOTIF (thus used in \TheName{}) is the same as EMBER, and feature extraction is conducted through the LIEF project~\cite{LIEF}, a lightweight PE parser. In general, the static features used are not engineered for detecting certain families but rather provide a comprehensive summary of a PE file. We elided certain feature extraction and pre-processing steps as they can be found in detail in the EMBER dataset~\cite{anderson2018ember}. 
Below, we discuss pre-processing steps and subsets of BODMAS and MOTIF we use for particular experiments. 

\begin{table}[tbp]
    \centering
    \setlength{\tabcolsep}{1mm}{
    \caption{Performance of \TheName{} with 1\% labeled samples, evaluated using varying numbers of BODMAS families. \label{tab:motivation}}
    \scalebox{0.96}{\begin{tabular}{l c c c c c c} 
        \toprule
        \# Families & $20$ & $30$ & $40$ & $60$ & $80$ & $100$\\ 
        \midrule
        Accuracy (\%) & 83.35 & 82.89 & 75.06 & 74.88 & 74.99 & 70.65 \\
        \bottomrule
    \end{tabular}}
    }
\end{table}

\subsection{Motivation for dataset pre-processing}\label{sub:motivation-dataset-pre-processing}
We train and evaluate \TheName{} on subsets of BODMAS and MOTIF consisting of all samples in the top 20 to top 50 most prevalent malware families.
We justify these choices as follows: 
First, as shown in Table~\ref{tab:motivation}, \TheName{} can perform multi-way classification with a large number of classes.
We further provide two additional experiments in Section~\ref{sec:appendix:all} in the Appendix to demonstrate \TheName{}'s effectiveness on the entire BODMAS dataset. In BODMAS, the top 20 to 30 most prevalent families account for $69.42\%$ to $80.02\%$  of all malware samples. In MOTIF, the top 50 most prevalent families account for $50.50\%$ of all samples.
Therefore, selecting these top families strikes a balance between covering a diverse subset of BODMAS/MOTIF and ensuring experimental expediency.
Second, most existing ML-based classifiers are typically evaluated on datasets with fewer than 25 classes~\cite{tekerek2022novel, catak2021data, fang2019semi, vasan2020image, chaganti2022image}.
By conducting 20-to-50-way classification, we can sufficiently demonstrate the effectiveness of our model in settings that follow current best practices.
Third, \TheName{}'s model design assumes that practitioners  have pre-identified families of interest and provided at least one labeled sample from each family.
In practice, if novel samples from an unseen family appear in the wild, \TheName{} requires only a small number of labeled examples of that new family to retrain the model, as demonstrated in Section~\ref{sec:rq4}.
We devise several datasets based on BODMAS and MOTIF to evaluate our classifier, each described below.

\begin{table}[tbp]
    \centering
    \setlength{\tabcolsep}{2mm}
    \caption{Dataset Statistics of BODMAS-20, MOTIF-50, \& BODMAS-Temporal.}\label{tab:overall_data}
    \scalebox{0.82}{\begin{tabular}{l l r r} 
        \toprule
        Training Design & Data Type & \# Class & \# Examples \\
        \midrule
        BODMAS-20 & Train & 20 & 31816 \\
         & Test & 20 & 7954 \\
         \midrule
        MOTIF-50 & Train & 50 & 1250 \\
         & Test & 50 & 313 \\
        \midrule
        BODMAS-Temporal & Train Pre-2020 & 20 & 8017 \\
         & Test Pre-2020 & 20 & 2005 \\
         & Test 2020 Q1 & 20 & 9808 \\
         & Test 2020 Q2 & 20 & 10075 \\
         & Test 2020 Q3 & 20 & 9865 \\
         & Test 2020 & 20 & 29748 \\
        \bottomrule
    \end{tabular}}
\end{table}

\begin{table}[tbp]
    \centering
    \setlength{\tabcolsep}{2mm}
    \caption{Dataset Statistics of BODMAS-LeaveOut containing three set-ups, each comprising two scenarios.}\label{tab:leave-out}
    \scalebox{0.80}{\begin{tabular}{ m{1.3cm} l m{1.1cm} m{1.1cm} m{1.4cm} m{2.0cm}}
        \toprule
        Set-up \newline (\# Class) & Scenario & \# Train \newline Examples & \# Test \newline Examples & \# Unident. Families    & \# Unident.-Fam. Examples in Test\\
        \midrule
          21  & Zero-shot & 32469 & 8118 & 1 &  413\\
          & One-shot & 32469 &  8118 & 0 & 0\\
          \midrule[0.01pt]
         25  & Zero-shot & 34688 & 8673 & 5 & 2699\\
           & One-shot & 34688 & 8673 & 0 & 0 \\
          \midrule[0.01pt]
          30  & Zero-shot & 36677 & 9170 & 10 & 4090 \\
           & One-shot & 36677 & 9170 & 0 & 0 \\
        \bottomrule
    \end{tabular}}
\end{table}
\noindent\textbf{BODMAS-20 and MOTIF-50:} We curated BODMAS-20 and MOTIF-50 to evaluate the performance of \TheName{} in standardized semi-supervised classification settings. 
BODMAS-20 consists of all samples within the 20 most prevalent families in BODMAS, and MOTIF-50 consists of all samples within the 50 most prevalent families in MOTIF. BODMAS-20 is relatively balanced with the largest family, \texttt{sfone}, consisting of 4729 samples and the smallest family, \texttt{unruy}, consists of 865 samples. Within MOTIF-50, the family sizes range from 142 to 16 samples. BODMAS-20 and MOTIF-50 both use an $80\%/20\%$ train/test split. 
Additionally, since \TheName{} relies on retrieving non-interpolatable features, Appendix~\ref{sec:appendix:noninterpolatable-overlap} also provides dataset statistics on train-test overlap of non-interpolatable features.

\noindent\textbf{BODMAS-Temporal:} We curate BODMAS-Temporal to evaluate the performance of \TheName{} when all testing families are represented in the training set.
However, the training and testing splits are taken from non-overlapping points in time (based upon dates in BODMAS), thus potentially capturing \emph{concept drift} where distributions shift over time~\cite{yang2021bodmas}. 
All samples from the top 20 most prevalent families are included.
However, the training set only consists of samples dated before 2020.
Most of the test set consists of samples gathered at a later time from a distinct distribution~\cite{yang2021bodmas}.
Specifically, five test sets are created: timestamped before 2020 (for a baseline), each quarter during 2020 (Q1, Q2, and Q3), and all samples from 2020.
Each train and test set contains at least one sample from each of the 20 families.
This dataset uses a 80\%/20\% train/test split on samples dated before 2020.
Table~\ref{tab:overall_data} summarizes the statistics of BODMAS-20, MOTIF-50, and BODMAS-Temporal.

\noindent\textbf{BODMAS-LeaveOut:} BODMAS-LeaveOut is curated to evaluate \TheName{} in handling malware from previously unlabeled families.  
To this end, we consider three training set-ups: (1) the Top 21 most prevalent families are selected with one family deemed `unidentified' during training;
(2) Top 25 families with five `unidentified'; and
(3) Top 30 families with ten `unidentified'.
As illustrated by Table~\ref{tab:leave-out}, each of these three testing set-ups contains two scenarios: `Zero-shot' and `One-shot'.
In `Zero-shot,' families deemed `unidentified' are not labeled in training.
In `One-shot,' we introduce a single labeled instance for each originally `unidentified' family during training, so that all families become newly identified.
The `unidentified' families and the one-shot examples are selected randomly.
The train-test split is performed before the `unidentified' families are selected, so that samples from all families are present in both train and test sets.
Note that by choosing these set-ups, it remains constant that 20 families are always `identified,' and thus labeled in training.
This design enables the assessment of \TheName{} when faced with completely unseen malware families, as well as its capacity to be retrained to handle previously-unlabeled families for which only one sample is newly-labeled.


\section{Evaluation Design}

We consider four research questions:

\squishlist
    \item RQ1: How well does \TheName{} sustain satisfactory performance with decreasing labeled training data in semi-supervised malware classification?
    \item RQ2: Does \TheName{} outperform state-of-the-art, static-feature-based, semi-supervised learning models in few-shot settings? 
    \item RQ3: How effective is \TheName{} when classifying malware from different timeframes?
    \item RQ4: How effective is \TheName{} when classifying samples from families unidentified in training?
\squishend

In this section, we discuss the methodology for answering each RQ.  
Section~\ref{sec:analysis} presents the empirical results.

\noindent\textbf{RQ1: Saturation Analysis} First, we assess how well \TheName{} can maintain satisfactory performance with decreasing amounts of labeled data through saturation analysis. This analysis directly addresses a key challenge in classifying evolving malware: the trade-off between the time and human resources needed to create larger labeled datasets vs. immediate, satisfactory model accuracy. 

Using BODMAS-20, we provide comprehensive evaluation of this trade-off by randomly selecting $0.1\%$, $0.5\%$, $1\%$, $2\%$, $5\%$, $10\%$, $50\%$, and $75\%$ of all data in training set as labeled data, ensuring that at least one sample from each family is labeled. 
In the $0.1\%$ setting, no family contains more than 5 labeled samples.

\noindent\textbf{RQ2: Model Comparison} 
\TheName{} aims to enhance malware classification using static malware features, an area of significant research importance~\cite{abusitta2021malware, aghakhani2020malware}. Thus, we measure \TheName{}'s relative performance in standardized few-shot settings by comparing with existing static-feature-based semi-supervised learning models on BODMAS-20 and MOTIF-50.
We consider the performance of seven baseline semi-supervised methods:
Pseudo-Label~\cite{lee2013pseudo}, Virtual Adversarial Training (VAT)~\cite{miyato2018virtual}, Virtual Adversarial Training equipped with our proposed retrieval-augmented pseudo-malware data (VAT aug), MixUp~\cite{zhang2017mixup}, MixMatch without data augmentation (MixMatch)~\cite{berthelot2019mixmatch}, MixMatch applied with Gaussian-noise-injection data augmentation (MixMatch Gaussian), and MORSE~\cite{wu2023grim}.
These techniques encompass a swath of state-of-the-art techniques as baselines for comparison.
Specifically, Pseudo-Label uses entropy-minimization by iteratively creating hard labels on high-confidence predictions, which are then incorporated into the labeled set.
In contrast, VAT strategically perturbs the labeled data to maximize changes in output class distributions, making the model more robust against variations from the training set.
Although MixUp is traditionally considered a regularization method for supervised learning~\cite{berthelot2019mixmatch}, we adapt the model and empirically verify that it is well-suited for few-shot malware classification tasks and thus incorporate it as a baseline method.
Last, MixMatch is a widely-used image-based semi-supervised learning method. 
Its overarching framework aligns closely with \TheName{}, though \TheName{}'s augmentation approach innovates through more sensible feature manipulations via domain-knowledge-awareness. 
Given that the traditional image-based augmentation techniques are not directly applicable to malware, we adopt two variants of MixMatch: one without any augmentation and another with Gaussian noise injection. 
In the latter, the noise is random, drawing from the Gaussian distribution of each family in the labeled set. This set-up allows for a fair comparison with \TheName{}, which introduces novel augmentation strategies tailored for malware classification.
Finally, MORSE is a state-of-the-art malware family classification model originally designed to mitigate label noise and class imbalance through semi-supervised learning. Given its demonstrated effectiveness in label-scarce scenarios, its semi-supervised architecture, and its malware-specific data augmentation techniques, MORSE can be adapted to our few-shot setting with appropriate modifications. Refer to Appendix~\ref{sec:appendix:morse} for more details on MORSE and its adaptations.

When evaluating on BODMAS-20, we consider $0.1\%$ of training set as labeled data. On MOTIF-50, we consider $1\%$ of the training set as labeled. We ensure at least one sample from each family is labeled, and highlight that these settings are close to the one-shot regime.

\noindent\textbf{RQ3: Temporal Analysis} Our third task is to evaluate \TheName{}'s robustness against temporal shift of malware distribution through leveraging test samples  collected from different periods of time than the training samples. 
We use BODMAS-Temporal, selecting $1\%$ of the training set as labeled data because it yields a desired trade-off between the number of labeled samples and model accuracy (discussed in Section~\ref{sub:rq1}). We compare \TheName{}'s performance with the seven other baseline methods mentioned above.

\noindent\textbf{RQ4: Leave-Out Testing} 
Next, we evaluate \TheName{}'s ability to quickly retrain to handle test samples representing malware families not originally identified in training. Like many ML malware classifiers~\cite{wu2023grim, yang2021bodmas, someya2023fcgat}, \TheName{} relies on a closed-world assumption, whereby it can only classify samples into families labeled in training. In real-world settings, when a new malware batch arrives containing at least one previously unseen family, practitioners using closed-world models can first leverage existing methods (e.g.~\cite{barbero2022transcending}) to reject unfamiliar samples and flag them for manual labeling. Once new families are labeled, we can retrain the model under a closed-world framework.

\TheName{}'s few-shot ability allows it to be promptly retrained to classify new families as soon as one sample from each previously unidentified family is labeled. To simulate how effective \TheName{} handles this retraining, we employ BODMAS-LeaveOut, which features a mix of predominantly `identified' families alongside at least one `unidentified' family. For each family deemed `identified' in training, we randomly select $1\%$ of the training set as labeled data. In the `Zero-shot' settings, we perform a random train/test split on the whole batch, after which samples from all families are present in both train and test sets. However, at this point, the `unidentified' families in the training set remain completely unlabeled and cannot be used as helpful training data. This setting provides a baseline for how \TheName{} (without retraining) would have performed under the existence of unseen families, potentially misclassifying them into existing families. In contrast, the `One-shot' settings simulate the addition of a single labeled instance for each previously `unidentified' family, after which \TheName{} is retrained. The performance improvement observed between the ‘Zero-shot’ and ‘One-shot’ scenarios quantifies \TheName{}'s ability to correctly classify newly labeled malware samples with minimal labeling effort.

\section{Results}\label{sec:analysis}

In this section, we present an evaluation of \TheName{}. 
We discuss each research question through: 
(RQ1) a saturation analysis to demonstrate model performance with respect to the volume of labeled malware samples; 
(RQ2) a model comparison with state-of-the-art models and techniques;
(RQ3) an evaluation of model robustness when classifying new samples from a later period of time than samples used in training; and 
(RQ4) an evaluation of model retraining when classifying samples in families that were previously unlabeled.

\begin{table}[b]
    \centering
    \setlength{\tabcolsep}{1mm}{
    \caption{Saturation evaluation of \TheName{} against BODMAS-20, with Fully Connected ResNet and LightGBM provided as a supervised standard for reference.
    \label{tab:rq1}}
    \scalebox{0.86}{\begin{tabular}{l r c c c c}
        \toprule
        Models & Labels & Accuracy & Precision  & Recall & F1 \\
        \midrule
        \multirow{9}{*}{\TheName{}} & $0.1\%$ & 73.88$\pm$2.67 & 79.61$\pm$1.36 & 75.64$\pm$2.75 & 74.07$\pm$2.35 \\
         & $0.5\%$ & 80.31$\pm$1.84 & 83.56$\pm$0.85 & 83.08$\pm$2.67 & 81.43$\pm$2.37 \\ 
         & $1\%$ & 83.35$\pm$1.03 & 85.24$\pm$0.62 & 86.08$\pm$1.24 & 84.60$\pm$1.31 \\
         & $2\%$ & 84.62$\pm$0.44 & 86.15$\pm$0.70 & 87.42$\pm$0.88 & 85.93$\pm$0.48 \\
         & $5\%$ & 86.20$\pm$0.27 & 87.47$\pm$0.60 & 89.23$\pm$0.49 & 87.55$\pm$0.45 \\
         & $10\%$ & 87.76$\pm$0.38 & 88.87$\pm$0.31 & 90.78$\pm$0.28 & 89.07$\pm$0.33\\
         & $25\%$ & 88.82$\pm$0.22 & 90.24$\pm$0.54 & 91.67$\pm$0.24 & 90.09$\pm$0.17 \\
         & $50\%$ & 89.28$\pm$0.27 & 90.42$\pm$0.28 & 92.12$\pm$0.11 & 90.65$\pm$0.14 \\
         & $75\%$ & 89.39$\pm$0.23 & 90.61$\pm$0.20 & 92.26$\pm$0.26 & 90.84$\pm$0.28 \\
        \midrule
        FC-ResNet~\cite{he2016deep} & $100\%$ & 93.45 & 93.72 & 94.07 & 93.87 \\ 
        \midrule[0.01pt]
        LightGBM~\cite{ke2017lightgbm} & $100\%$ & 93.90 & 94.09 & 94.01 & 94.04 \\ 
        \bottomrule
    \end{tabular}}
    }
\end{table}

\begin{table*}[tbp] 
    \centering
    \setlength{\tabcolsep}{1mm}{
    \caption{Comparison of \TheName{} against semi-supervised methods on BODMAS-20 ($0.1\%$ labeled data) and MOTIF-50 ($1\%$ labeled data), where most families are identified by one-shot labels. 
    \label{tab:rq2}}
    \scalebox{0.86}{\begin{tabular}{l c c c c c c c c } 
        \toprule
        & \multicolumn{4}{c}{BODMAS-20} & \multicolumn{4}{c}{MOTIF-50}\\
        \cmidrule(lr){2-5} \cmidrule(lr){6-9}
        Models & Accuracy & Precision (macro) & Recall (macro) & F1 (macro) & Accuracy & Precision (macro) & Recall (macro) & F1 (macro)\\ 
        \midrule
        Pseudo-Label~\cite{lee2013pseudo} & 44.23$\pm$5.33 & 59.44$\pm$4.85 & 40.58$\pm$4.90 & 38.30$\pm$2.93 & 9.41$\pm$0.92 & 9.07$\pm$0.82 & 9.28$\pm$1.24 & 6.96$\pm$0.97\\
        VAT~\cite{miyato2018virtual} & 67.56$\pm$4.30 & 74.10$\pm$2.14 & 68.01$\pm$3.62 & 64.50$\pm$3.63 & 33.55$\pm$3.93 & 38.83$\pm$7.81 & 35.52$\pm$4.72 & 31.18$\pm$5.72\\
        VAT~(aug)~\cite{miyato2018virtual} & 71.77$\pm$6.01 & 77.16$\pm$2.92 & 74.02$\pm$2.85 & 72.38$\pm$2.28  & 36.81$\pm$4.94 & 41.74$\pm$5.71 & 38.15$\pm$5.69 & 35.32$\pm$6.30\\
        MixUp~\cite{zhang2017mixup} & 69.96$\pm$3.80 & 79.18$\pm$2.85 & 71.65$\pm$3.46 & 69.75$\pm$2.21  & 39.49$\pm$5.04 & 43.83$\pm$6.22 & 42.32$\pm$5.33 & 37.06$\pm$5.84\\
        MixMatch~\cite{berthelot2019mixmatch} & 69.85$\pm$7.48 & 79.47$\pm$1.54 & 69.94$\pm$5.20 & 67.96$\pm$4.28 & 36.77$\pm$6.84 & 47.32$\pm$6.57 & 37.88$\pm$6.51 & 35.01$\pm$6.13\\
        MixMatch~(Gaussian)~\cite{berthelot2019mixmatch} & 68.72$\pm$5.60 & 79.37$\pm$2.65 & 67.42$\pm$6.03 & 66.53$\pm$4.72 & 34.89$\pm$6.47 & 50.56$\pm$8.01 & 35.61$\pm$6.44 & 35.81$\pm$6.44\\
        \midrule[0.01pt]
        MORSE~\cite{wu2023grim} & 70.09$\pm$3.83 & 77.60$\pm$2.10 & 70.51$\pm$2.16 & 67.95$\pm$2.61 & 38.59$\pm$4.78 & 47.97$\pm$9.02 & 40.56$\pm$5.87 & 36.25$\pm$5.90\\
        \midrule
        \TheName{} & \textbf{73.88$\pm$2.67} & \textbf{79.61$\pm$1.36} & \textbf{75.64$\pm$2.75} & \textbf{74.07$\pm$2.35} & \textbf{40.45$\pm$4.77} & 45.45$\pm$6.59 & \textbf{43.06$\pm$5.31} & \textbf{39.70$\pm$5.16}\\
        \bottomrule
    \end{tabular}}
    }
\end{table*}

\subsection{RQ1: Saturation Analysis}\label{sub:rq1}
In this subsection, we measure how well \TheName{} retains adequate performance when its data augmentation pipelines is provided with varying number of labeled data for reference.
We use a standardized semi-supervised set-up 
over the BODMAS-20 dataset, and report accuracy, precision, recall, and F1 (all macro) as key criteria.

\noindent\textbf{Result for Saturation Analysis} Table~\ref{tab:rq1}  
shows that substantially decreasing the volume of labeled samples does not dramatically undermine the performance of \TheName{} on BODMAS-20.
We found a labeled set consisting of $1\%$ of all data  balances the number of labeled samples with overall model performance on this dataset. Furthermore, \TheName{} seems unbiased across a variety of families from different paradigms, as shown in Table~\ref{tab:group} in the Appendix.
We note that \TheName{}'s pervasive use of regularization likely means the model is underfit and thus cannot reach the performance of a fully-supervised model (e.g., Fully Connected ResNet), which is the gold standard for classification and serves as a reference.
Nonetheless, our approach can provide adequate classification accuracy with very few labeled samples.



Furthermore, the supervised performance of our chosen base classifier, FC-ResNet (Appendix~\ref{sec:appendix:implementation}), as shown in Table~\ref{tab:rq1}, demonstrates that FC-ResNet possesses sufficient expressiveness to serve as the base classifier within our semi-supervised framework. Its performance remains competitive with LightGBM~\cite{ke2017lightgbm}, a state-of-the-art model used in the original BODMAS experiments~\cite{yang2021bodmas}. See more discussion on FC-ResNet's expresiveness in Appendix~\ref{sec:appendix:resnet}.

\subsection{RQ2: Model Comparison}\label{sec:comparison}

We further evaluate \TheName{} in near one-shot settings on BODMAS-20 and MOTIF-50 by comparing against seven other baseline models trained on static features: Pseudo-Label, VAT, VAT (aug), MixUp, MixMatch, MixMatch (Gaussian), and MORSE. 
We use accuracy, precision, recall, and F1 (all macro) as criteria.

\noindent\textbf{Result for Model Comparison} Table~\ref{tab:rq2} compares \TheName{} against baseline models with $0.1\%$ labeled data on BODMAS-20 and $1\%$ labels on MOTIF-50, where most families are identified by one-shot labels. 

On BODMAS-20, \TheName{} achieves a relative accuracy improvement of at least $5.41\%$ compared to all but one of the baseline models. The only model not outperformed by such a margin is VAT (aug), which employs our malware data augmentation method. 
In fact, VAT (aug), on either BODMAS-20 or MOTIF-50, sees a relative improvement of $6.23\%$ or $7.49\%$ over VAT. This demonstrates our augmentation technique also effectively enhances other compatible semi-supervised frameworks. 
Moreover, \TheName{} surpasses MixMatch with Gaussian noise data augmentation by a relative margin of $7.51\%$ on BODMAS-20 and $15.94\%$ on MOTIF-50, setting our proposed augmentation pipeline apart from semantics-agnostic, statistical augmentation approaches. 

\TheName{} outperforms MORSE, likely due to fundamental differences in design objectives. MORSE is tailored for scenarios with more than $10\%$ labeled data, making many of its elegant mechanisms impractical in few-shot settings. Furthermore, MORSE's proposed data augmentation, although malware-specific, still largely disregards the underlying semantics behind features. We provide further discussion on these points in Appendix~\ref{sec:appendix:morse}.

Figure~\ref{fig:BODMAS-few-shot-results} illustrates that the accuracy of \TheName{} and the baseline models converge as the number of labeled data increases. We note that all models perform significantly worse on MOTIF-50 due to the intrinsic complexities, noisy labels~\cite{wu2023grim}, and small sizes of malware classes in MOTIF.

On BODMAS-20 and MOTIF-50, \TheName{} mainly outperforms other models in Accuracy and F1 macro. This indicates that our classifier excels at identifying a larger portion of malware families overall and maintains a superior balance between minimizing false positives and negatives.

\begin{figure}[tbp]
\centering
\includegraphics[width=0.42\textwidth]{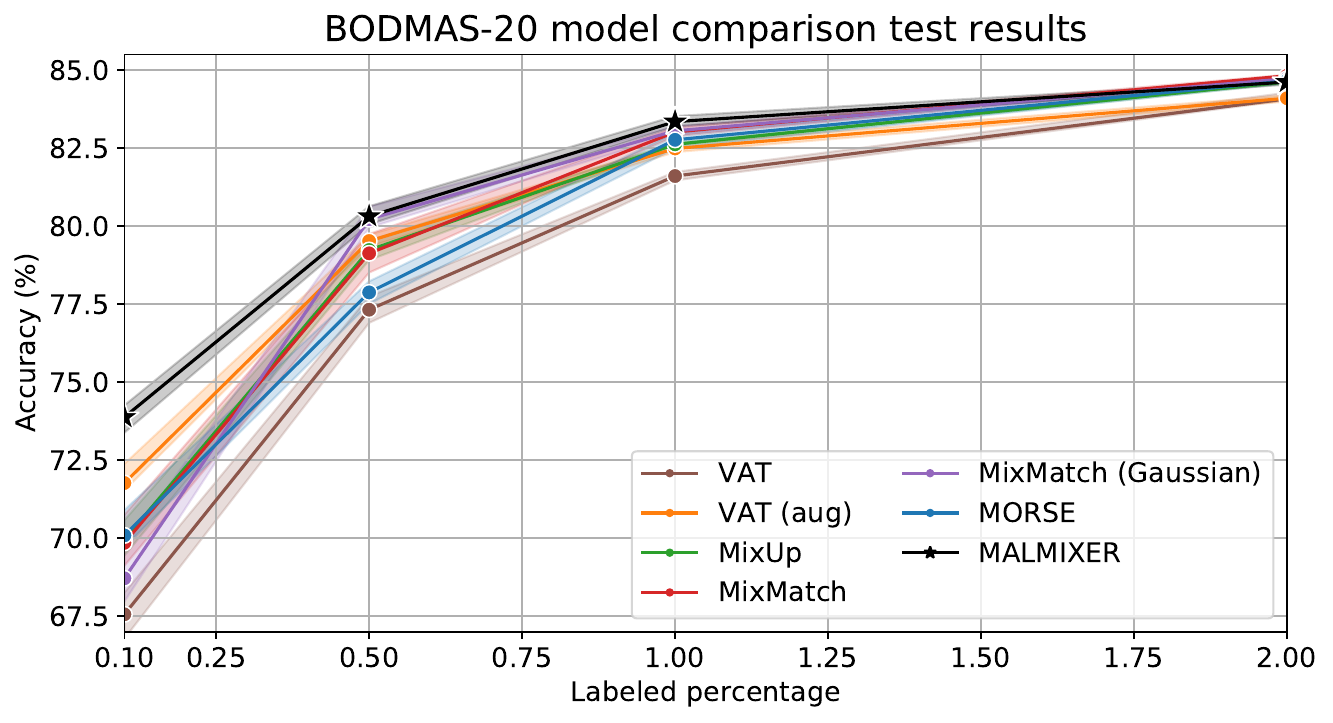}
\vspace{-8pt}
\caption{Comparison of \TheName{} to baseline methods in few-shot settings on BODMAS-20. \TheName{} outperforms all models with fewer (i.e. towards the leftmost end of the x-axis)  labeled data, but performance converges with a higher proportion of labeled data. 
}
\label{fig:BODMAS-few-shot-results}
\end{figure}

\begin{table*}[tbp]
    \centering
    \setlength{\tabcolsep}{1mm}{
    \caption{Comparison of \TheName{} against baseline methods on malware classification of BODMAS-Temporal.
    \label{tab:rq3}}
    \scalebox{0.84}{\begin{tabular}{l | r r r r r | r r r r r}
        \toprule
        & \multicolumn{5}{c|}{Accuracy} & \multicolumn{5}{c}{Precision (macro)} \\
        \midrule
        Models & Pre-2020 & 2020 Q1 & 2020 Q2 & 2020 Q3 & 2020 & Pre-2020 & 2020 Q1 & 2020 Q2 & 2020 Q3 & 2020 \\
        \midrule
        Pseudo-Label~\cite{lee2013pseudo} & 48.96$\pm$4.21 & 33.16$\pm$3.78 & 26.65$\pm$~6.89 & 21.20$\pm$6.94 & 27.01$\pm$11.87 & 58.81$\pm$6.69 & 53.40$\pm$4.36 & 45.78$\pm$2.25 & 39.76$\pm$6.72 & 46.32$\pm$8.91 \\
        VAT~\cite{miyato2018virtual} & 73.33$\pm$2.15 & 62.37$\pm$2.14 & 50.45$\pm$~3.09 & 53.80$\pm$5.23 & 55.54$\pm$~9.60 & 66.20$\pm$4.25 & 68.09$\pm$3.49 & 62.86$\pm$4.80 & 63.34$\pm$6.97 & 64.76$\pm$5.12\\
        VAT~(aug)~\cite{miyato2018virtual} & 76.70$\pm$2.92 & 66.86$\pm$3.28 & 60.29$\pm$11.16 & 63.36$\pm$5.87 & 63.50$\pm$~8.81 & 74.91$\pm$3.97 & 71.71$\pm$3.78 & 69.06$\pm$4.42 & 66.64$\pm$5.14 & 69.14$\pm$5.09 \\
        MixUp~\cite{zhang2017mixup} & 78.52$\pm$2.40 & 67.62$\pm$2.84 & 55.51$\pm$~9.83 & 60.88$\pm$6.08 & 61.34$\pm$10.43 & 76.73$\pm$4.24 & 76.15$\pm$2.51 & 72.37$\pm$2.44 & 71.73$\pm$2.54 & 73.42$\pm$3.59 \\
        MixMatch~\cite{berthelot2019mixmatch} & 77.60$\pm$2.32 & 68.00$\pm$1.67 & 63.73$\pm$~9.44 & 62.04$\pm$7.32 & 64.59$\pm$~8.37 & 74.74$\pm$3.04 & 75.71$\pm$2.71 & 72.27$\pm$3.81 & 73.22$\pm$3.11 & 73.73$\pm$3.23 \\
        MixMatch~(Gaussian)~\cite{berthelot2019mixmatch} & 76.96$\pm$1.86 & 67.57$\pm$3.21 & 65.16$\pm$~6.00 & 61.54$\pm$4.61 & 64.76$\pm$~7.01 & 74.52$\pm$3.65 & 75.18$\pm$3.23 & 71.15$\pm$2.30 & 72.38$\pm$2.83 & 72.90$\pm$3.25 \\
        \midrule[0.01pt]
        MORSE~\cite{wu2023grim} & 77.40$\pm$1.57 & 62.22$\pm$1.85 & 49.41$\pm$~1.19 & 58.13$\pm$4.20 & 57.59$\pm$~7.16 & 74.68$\pm$4.12 & 74.54$\pm$2.26 & 67.53$\pm$3.98 & 69.22$\pm$5.73 & 70.43$\pm$4.99 \\
        \midrule
        \TheName{} & \textbf{79.52$\pm$1.56} & \textbf{71.75$\pm$2.99} & \textbf{72.31$\pm$~3.62} & \textbf{72.45$\pm$5.36} & \textbf{72.17$\pm$~4.68} & \textbf{78.51$\pm$2.51} & \textbf{76.22$\pm$2.79} & \textbf{74.56$\pm$1.74} & \textbf{76.06$\pm$2.07} & \textbf{75.61$\pm$2.57} \\
        \toprule
        & \multicolumn{5}{c|}{Recall (macro)} & \multicolumn{5}{c}{F1 (macro)} \\
        \midrule
        Models & Pre-2020 & 2020 Q1 & 2020 Q2 & 2020 Q3 & 2020 & Pre-2020 & 2020 Q1 & 2020 Q2 & 2020 Q3 & 2020\\
        \midrule
        Pseudo-Label~\cite{lee2013pseudo} & 30.85$\pm$4.09 & 27.11$\pm$2.12 & 26.07$\pm$3.36 & 24.22$\pm$3.43 & 25.80$\pm$3.94 & 31.68$\pm$4.29 & 25.39$\pm$2.48 & 23.02$\pm$4.05 & 18.35$\pm$3.06 & 22.25$\pm$5.91 \\
        VAT~\cite{miyato2018virtual} & 66.30$\pm$3.82 & 59.25$\pm$3.48 & 52.83$\pm$1.09 & 54.07$\pm$3.62 & 55.38$\pm$6.19 & 61.93$\pm$2.25 & 55.46$\pm$3.10 & 50.03$\pm$1.03 & 48.66$\pm$2.75 & 51.38$\pm$5.80 \\
        VAT~(aug)~\cite{miyato2018virtual} & 74.75$\pm$2.75 & 65.52$\pm$1.95 & 61.52$\pm$3.05 & 60.57$\pm$4.26 & 62.54$\pm$6.42 & 71.14$\pm$3.47 & 62.86$\pm$1.74 & 58.79$\pm$4.79 & 54.91$\pm$4.03 & 58.85$\pm$7.04 \\
        MixUp~\cite{zhang2017mixup} & 72.81$\pm$5.97 & 64.41$\pm$2.66 & 58.19$\pm$3.19 & 60.04$\pm$3.87 & 60.88$\pm$6.91 & 70.35$\pm$5.15 & 62.73$\pm$2.27 & 56.86$\pm$3.56 & 53.69$\pm$4.89 & 57.76$\pm$7.53 \\
        MixMatch~\cite{berthelot2019mixmatch} & 72.45$\pm$4.28 & 64.26$\pm$1.30 & 58.16$\pm$1.92 & 60.39$\pm$3.70 & 60.94$\pm$6.25 & 69.41$\pm$3.99 & 61.87$\pm$1.30 & 56.70$\pm$2.95 & 55.18$\pm$3.97 & 57.91$\pm$6.44 \\
        MixMatch~(Gaussian)~\cite{berthelot2019mixmatch} & 73.77$\pm$3.01 & 64.59$\pm$1.19 & 58.82$\pm$0.68 & 60.19$\pm$2.18 & 61.20$\pm$6.27 & 70.51$\pm$2.61 & 62.25$\pm$1.98 & 56.57$\pm$2.33 & 55.15$\pm$2.23 & 57.99$\pm$6.54 \\
        \midrule[0.01pt]
        MORSE~\cite{wu2023grim} & 72.77$\pm$2.86 & 62.13$\pm$1.81 & 54.28$\pm$2.37 & 59.03$\pm$4.41 & 58.48$\pm$4.39 & 69.59$\pm$2.91 & 59.73$\pm$1.31 & 50.75$\pm$2.97 & 51.12$\pm$5.97 & 53.86$\pm$5.59 \\
        \midrule
        \textbf{\TheName{}} & \textbf{79.67$\pm$3.83} & \textbf{70.58$\pm$2.64} & \textbf{65.74$\pm$3.23} & \textbf{66.52$\pm$4.74} & \textbf{67.61$\pm$6.61} & \textbf{74.71$\pm$2.18} & \textbf{67.43$\pm$1.12} & \textbf{63.70$\pm$3.28} & \textbf{63.58$\pm$3.48} & \textbf{64.90$\pm$5.25} \\
        \bottomrule
    \end{tabular}}
    }
\end{table*}

\begin{tcolorbox}[colback=white, colframe=black, colbacktitle=white, coltitle=black, boxsep=0pt]
The few-shot performance of \TheName{} in results for RQ1--2 showcases our augmentation approach's effectiveness in overcoming small labeled datasets.
\end{tcolorbox}

\begin{figure}[tbp]
\centering
\includegraphics[width=0.42\textwidth]{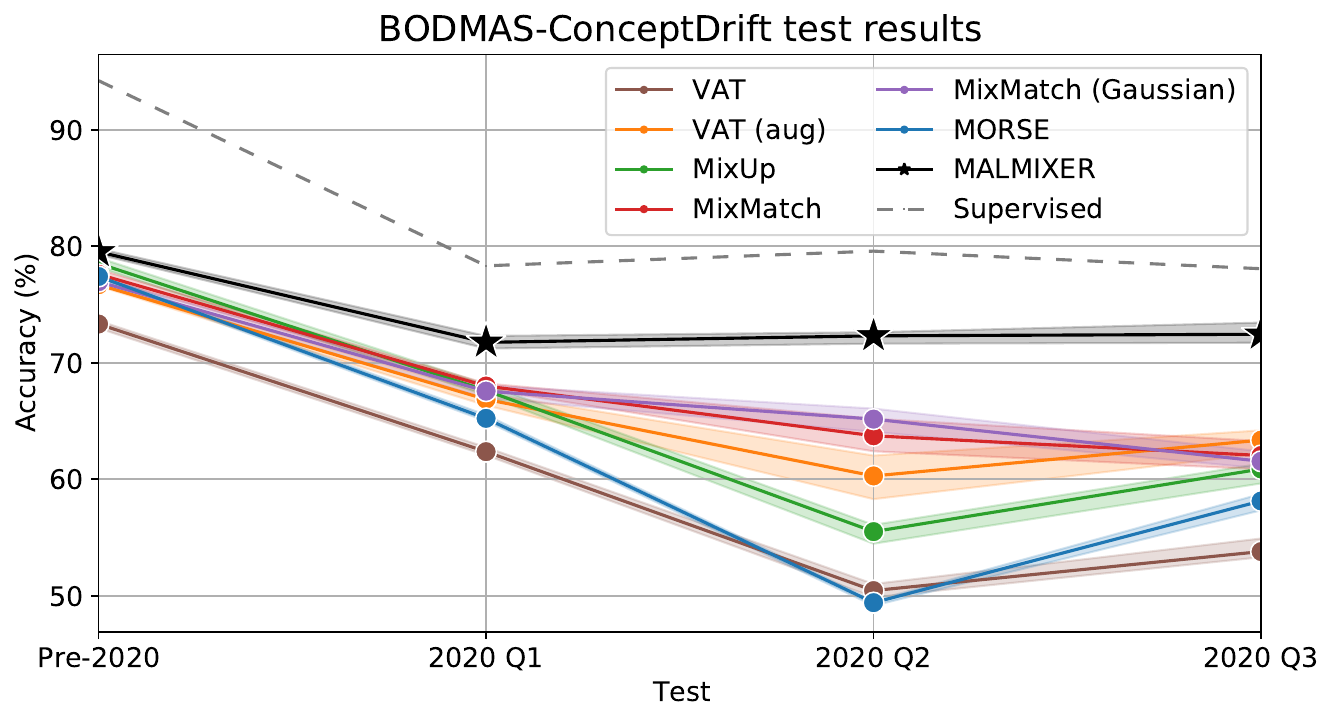}
\vspace{-8pt}
\caption{Accuracy comparison of \TheName{} to baseline methods on BODMAS-Temporal test sets. 
}\label{fig:BODMAS-ConceptDrift-results}
\end{figure}

\subsection{RQ3: Temporal Analysis}
Next, we assess \TheName{} in a temporal analysis to analyze \TheName{}'s ability to handle new samples from previously identified families.  
In practice, new malware samples may evolve over time --- polymorphic or metamorphic malware may exhibit similar behavior but appear drastically different. 
Thus, RQ3 is intended to explore whether our augmentation pipeline can help malware classifiers generalize to new samples from existing families.

In this setting, all malware families of interest are represented in train set, but the evaluation samples follow a different distribution of families.
We evaluate our model against the BODMAS-Temporal dataset, and report accuracy, precision, recall, and F1 (all macro) as key criteria.

\noindent\textbf{Result for Temporal Analysis} Table~\ref{tab:rq3} shows that \TheName{} surpasses all other models in both the in-domain test set (pre-2020) and temporally shifted test sets (2020 Q1/Q2/Q3, and all 2020). As shown in Figure~\ref{fig:BODMAS-ConceptDrift-results}, although all baseline models exhibit a decline in performance over time, indicating the impact of domain shift, \TheName{} maintains stable performance across all temporally shifted test sets. Remarkably, in 2020 Q3, our model outperforms all other models by a relative improvement of $11.44\%$ and is only $7.56\%$ below a fully supervised model in relative terms.

Our model's consistent performance across varying data distributions likely stems from its ability to synthesize meaningful malware features that mimic variation techniques that can alter features like entropy, file size, and import table while retaining the samples' overall semantics.
In contrast, although data perturbations created by other baseline models may enhance robustness, they may not effectively predict meaningful unseen samples mutated from observed samples.

\begin{tcolorbox}[colback=white, colframe=black, colbacktitle=white, coltitle=black, boxsep=0pt]
\indent \TheName{} effectively improves  classification of new samples from previously identified families.  
Our proposed data augmentation helps to generalize classifiers in the face of shifting distribution of in-the-wild malware over time.

\end{tcolorbox}

\subsection{RQ4: Leave-Out Analysis}\label{sec:rq4}
Next, we consider a scenario in which new, previously unlabeled malware families must be classified.
In practice, new malware may appear over time as attackers find new zero-day exploits. 
Classifying these unseen-family instances using \TheName{} requires retraining the model after experts manually provide labels for each new family.
Using the BODMAS-LeaveOut dataset, we evaluate the performance improvement of \TheName{} when as few as one labeled sample from each previously `unidentified' family is introduced to training. 
We employ accuracy, precision, recall, and F1 (all macro) as criteria.

\textbf{Result for Leave-Out Analysis}
Table~\ref{tab:realWorld} illustrates improvements in \TheName{}'s performance when one-shot instances from formerly `unidentified' families become labeled.
In the set-up with 21 malware families, the introduction of merely one additional label for the originally `unidentified' malware family (constituting about $5\%$ of all data) increases \TheName{}'s accuracy by $4.57\%$ through straightforward retraining, which can be completed within an hour.
In more challenging scenarios, where 10 families (accounting for roughly $44\%$ of all data) in the test set are deemed `unidentified' in training, 
the addition of just 10 labels can increase the model's accuracy by a substantial $21.53\%$.

\begin{tcolorbox}[colback=white, colframe=black, colbacktitle=white, coltitle=black, boxsep=0pt]
\indent Leave-out results imply \TheName{}'s performance improves effectively through a quick retraining with minimal labeling of previously unidentified families, reducing the amount of manual labeling required to classify a new family.
\end{tcolorbox}

\subsection{Ablation Studies}\label{sec:appendix:ablation}

Last, we conduct ablation studies to understand how each component of our data augmentation pipeline contributes to \TheName{}'s performance.
We evaluate this using the BODMAS-Temporal set-up, maintaining the training set while using  all 2020 data for testing.
Table~\ref{tab:ablation} displays the performance decline under three conditions: (1) without the alignment-based malware selection step, where we retrieve any real set of non-interpolatable features representative of the mixed instance, disregarding its compatibility with the mutated interpolatable features;
(2) without both the retrieval-based malware matching and alignment-based malware selection steps, where we eliminate the use of embeddings altogether and perform data augmentation by directly interpolating two malware instances' original features without semantics awareness; and 
(3) without our mixing-based data augmentation, where we solely rely on Gaussian noise augmentation.

 \begin{table}[tbp]
    \centering
    \setlength{\tabcolsep}{1mm}{
    \caption{\TheName{}   on BODMAS-LeaveOut when one-shot labeled instances are introduced to previously unidentified families.
    \label{tab:realWorld}}
    \scalebox{0.84}{\begin{tabular}{m{1.1cm} l c c c c}
        \toprule
        Set-up \newline (\# Class) & Scenario & Accuracy & Precision & Recall & F1 \\
        \midrule
        21 & Zero-shot & 76.81$\pm$1.02 & 77.22$\pm$0.81 &80.17$\pm$2.09 & 76.59$\pm$2.25 \\
        & \textbf{One-shot} & \textbf{81.38$\pm$1.09} & \textbf{83.79$\pm$1.12} & \textbf{84.65$\pm$2.20} & \textbf{82.37$\pm$2.43}  \\ 
        \midrule[0.01pt]
         25 & Zero-shot & 60.38$\pm$1.18 & 60.09$\pm$1.28 & 69.33$\pm$1.64 & 61.35$\pm$1.42 \\
         & \textbf{One-shot} & \textbf{71.84$\pm$4.27} & \textbf{78.95$\pm$3.96}
 & \textbf{78.64$\pm$3.18} & \textbf{74.49$\pm$3.72} \\
         \midrule[0.01pt]
        30 & Zero-shot & 49.25$\pm$1.18 & 46.62$\pm$1.91 & 57.69$\pm$2.45 & 47.50$\pm$2.04 \\
        & \textbf{One-shot} & \textbf{70.78$\pm$3.80} & \textbf{75.36$\pm$2.76}
 & \textbf{77.58$\pm$2.47} & \textbf{72.39$\pm$3.20} \\
        \bottomrule
    \end{tabular}}
    }
\end{table}

\begin{table}[tbp]
    \centering
    \setlength{\tabcolsep}{3mm}{
    \caption{Ablation results. 
    \label{tab:ablation}}
    \scalebox{0.86}{\begin{tabular}{l c }
        \toprule
        Ablation & Accuracy \\
        \midrule
        \TheName{} &  $72.17\pm4.68$\\
         \TheName{} without alignment-based selection &  $71.59\pm4.46$\\
         \TheName{} with direct instance-mixing augmentation  &  $70.81\pm5.87$\\
         \TheName{} with only Gaussian noise augmentation & $64.76\pm7.01$ \\
        \bottomrule
    \end{tabular}}
    }
\end{table}

\section{Discussion}\label{sec:discussion}
This section discusses \TheName{}'s implications for security and AI communities, as well as its limitations. 

\noindent\textbf{Interpretation and Implication of Findings} 
The collective findings from RQ1--4 suggest that \TheName{} can effectively improve few-shot malware family classification performance given an influx of new malware samples by being able to (1) accurately classify malware families with limited labels; (2) generalize to evolved instances from previously identified families without substantial new labeling; and (3) effectively retrain to detect new families with minimal new labels.
These findings imply that, in contrast to the semantics-agnostic image-based data augmentation, malware data augmentation pipelines can benefit from domain-knowledge-awareness achieved through augmenting different sets of malware features differently, as well as retrieval and alignment techniques capturing feature semantics. Furthermore, equipped with such domain-knowledge-aware data augmentation approaches, semi-supervised frameworks analogous to those implemented in \TheName{} can be successfully extended to malware family classification settings.
Intuitively, \TheName{} works because few-shot samples sparse within the true distribution of malware lead to difficulty drawing decision boundaries. Our approach uses domain knowledge to more faithfully mimic true malware family feature distribution, narrowing down decision boundaries.
In practice, we speculate that \TheName{} can help alleviate the extensive manual labor required to classify a large dataset.

\noindent\textbf{Limitations and Future Work}: We note that while our data augmentation approach is domain-knowledge-aware, it does not guarantee that augmented pseudo-samples can be reverse-engineered into an exact real-world malware executable.
That said, data augmentation methods that fully preserve the semantics of a malware executable often require semantics-aware transformations on malware binaries~\cite{wong2022marvolo}, or possibly another diffusion step to recover a real binary out of our augmented feature representations, both of which can be time and resource-intensive.
In our approach, we made a deliberate trade-off between some semantic validity and operational efficiency.
We contain our claim of ``semantic-awareness''
to the use of domain knowledge in order to become aware of what feature interpolations more/less likely change the underlying malware family semantics.
This is an improvement over conventional image-based data augmentations (which flip random bits without understanding any consequences), and has been shown effective with improved downstream ML classifiers' performance (Section~\ref{sec:analysis}) and the fact that augmented feature representations fall back into the clusters of ground-truth families (Section~\ref{sec:case}).
From a downstream ML model's perspective, our data augmentation approach, although not perfectly semantics-preserving, already better mimics the underlying distribution of each family in the feature space.

Currently, \TheName{} is evaluated only on BODMAS and MOTIF, both within the Windows PE malware domain. Future work could extend its applicability to other domains (e.g. Android malware), enhancing generalizability. In principle, adapting our approach to another domain involves manually re-defining the operations in Appendix~\ref{sec:appendix:interpolatable} (which took approximately five hours to complete by one author familiar with malware classification). 

Moreover, \TheName{} currently accepts malware instances represented by features obtained through static analysis. 
While static-feature-based classification is a crucial research domain~\cite{aghakhani2020malware} and is often more effective than end-to-end classification that consumes entire malware binaries~\cite{anderson2018ember}, future work can investigate efficient and semantics-aware augmentation methods that target dynamic analysis features. 
Nonetheless, the benefits of static analysis features, such as efficiency, scalability, and safety, hold substantial value for numerous practical applications, and are readily extracted from binaries. 

Additionally, static features are susceptible to code obfuscation. To the end, Appendix~\ref{sec:appendix:obfuscation} also evaluates \TheName{} on packed malware samples. While this experiment is limited by noisy family labels, it demonstrates negative impacts of packing on \TheName{}'s performance, likely due to the weakening of our $k$-NN and clustering assumptions in Sections~\ref{sec:intro} and \ref{sec:results-rq0}. Nevertheless, \TheName{} remains among the top-performing methods, and our semantics-aware data augmentation pipeline is at least competitive with Gaussian-noise-based, image-inspired data augmentation techniques under obfuscated conditions. These findings highlight the need for future research to enhance the robustness of static feature-based classifiers against obfuscation.

We also limited our saturation analysis to BODMAS-20. 
Appendix~\ref{sec:appendix:saturation} conducts an additional saturation analysis on MOTIF-50. Results show that \TheName{} achieves satisfactory performance in few-shot settings on MOTIF-50, although such satisfactory performance requires a higher percentage of labeled data compared to BODMAS-20. We hypothesize that the high label noise in MOTIF-50 renders it less suitable for an accurate saturation analysis, and thus omitted it from the main text.
Future work aimed at benchmarking models' performance degradation with decreasing numbers of labels via saturation analysis could benefit from additional Windows PE datasets featuring more carefully curated family labels.
In addition, we only evaluated \TheName{} on downsampled versions of BODMAS and MOTIF, focusing primarily on the top-N malware families. \TheName{} is designed to expedite classification for large families.
In real-world settings, the influx of malware data is typically characterized by a long-tail distribution with many small families. As experts generally have the capacity to promptly label only the larger, more prevalent families, the labeled pool naturally skews toward these major families. Our evaluation aligns the unlabeled data distribution and test distribution with that of the labeled samples, thus only evaluating \TheName{}'s performance on on large families. 
Appendix~\ref{sec:appendix:all} further evaluates \TheName{}'s performance on non-downsized datasets, in presence of more and smaller families. Results show that \TheName{}'s performance on large families remains strong, even with small unlabeled families falsely classified into these larger families. 
When only few-shot labels for large families are available, preliminarily misclassifying unlabeled small families into larger families can still provide insightful information about the characteristics of the unlabeled samples and allow engineers to prioritize further analysis. 
Appendix~\ref{sec:appendix:all} also shows \TheName{}'s performance naturally degrades with more families --- a well-known challenge in multi-way classification, as also observed with the supervised baseline. We acknowledge this limitation of our work, and leave enhancing robustness against an increasing number of classes as important future research.

\noindent\textbf{Practical Implications}
Our model operates on a closed-wold assumption, whereby it can only classify malware samples into families identified during training. This assumption aligns with real-world scenarios where we wish to train a classifier from scratch given an influx of unlabeled dataset. Once experts provide few-shot labels, likely for large families, \TheName{}'s few-shot capacity allows for prompt classifications of a largely unlabeled dataset, as discussed above.

In some other cases, we may wish to apply a pre-trained \TheName{} to a newly arrived test set containing malware families not seen during training, resulting in an open-world setting. \TheName{} is not designed or trained for a reject option for potentially unseen families. Instead, existing techniques~\cite{barbero2022transcending} (e.g. fully-supervised models explicitly designed and trained with decision threshold) can detect unknown families, almost surely outperforming few-shot \TheName{} on the rejection task. Once existing techniques reject new families and experts manually identify one labeled instance per new family, the problem space becomes closed-world, and \TheName{} becomes practically useful for what happens next. This scenario is simulated in Section~\ref{sec:rq4}, where \TheName{}’s few-shot capability allows it to quickly retrain to accurately classify all instances of these new families using few labeled samples, thereby reducing the number of labeled samples needed to understand an emerging family.

\section{Conclusion}\label{sec:conclusion}

We present \TheName{}, a few-shot malware family classifier unifying a novel, lightweight, domain-knowledge-aware malware data augmentation approach and a semi-supervised framework previously unexplored in malware contexts. Our experiments show that our classifier outperforms others in few-shot malware classification settings. 
This research verifies the feasibility and effectiveness of lightweight, domain-knowledge-aware data augmentation methods for malware features, which prove valuable when few labeled samples are available. 
We further showcase the capabilities of the implemented semi-supervised framework to bolster the effect of domain-knowledge-aware malware data augmentation and thereby improve few-shot malware family classification performance. 



\section*{Acknowledgment}

We acknowledge partial support from the NSA under grant H98230-23-C-0279, the NSF under grant DGE2312057, and an Amazon Research Award.  The views expressed here do not necessarily reflect those of the US Government.



%



\bibliographystyle{IEEEtran}
\bibliography{IEEEtran}

\newpage

\begin{appendices}

\section{Implementation Details} \label{sec:appendix:implementation}

We use two Nvidia A30 GPUs during model training and evaluation.
We set the random seed as 17 in all processes for reproducibility.
We note that training and evaluation across all experimental settings required less than one hour each using BODMAS-20 with $1\%$ labeled data.
In our experiments, we standardize all features in the BODMAS dataset through z-score scaling as in standard practice. 
For the encoder-decoder architecture, encoder $\phi_i$ comprises three layers, with each layer outputting 512 features. The corresponding decoder $\psi_i$ consists of three layers, mirroring the structure of the encoder in reverse order. Encoder $\phi_n$ has five layers, outputting 1024, 1024, 512, 512, and 512 features respectively. The corresponding the decoder $\psi_n$ reflects the encoder's architecture in reverse order. Consequently, the embeddings $h_u^d, h_u^s, h_i^d, h_i^s$ created by the encoder-decoder architecture are all represented as 256-dimensional vectors. 
For our retrieval-based data augmentation, we use Faiss~\cite{johnson2019billion}, an open-source library designed for scalable similarity search and clustering, to construct the k-nearest-neighbor graph and perform similarity searches with L2 distance.
Top 5 neighbors are selected for the $k$-NN graph in the original feature space $\mathcal{S}$ and 5 candidates are selected for alignment-based malware selection.

For the semi-supervised framework, unless otherwise noted, we use a Fully Connected ResNet (FC-ResNet)~\cite{he2016deep} as the base classifier for \TheName{} and all other semi-supervised methods evaluated. Our FC-Resnet model begins with a fully connected layer projecting the 2,381-dimensional input to a 1,024-dimensional latent space, followed by 12 residual blocks organized into four groups with progressively decreasing dimensions ($1,024-512-256-128$), where each group contains three residual blocks. Each block consists of two linear layers, LeakyReLU activations, and skip connections to aid gradient flow. Downsampling is applied when dimensions change. The final representation is flattened and passed through a fully connected classification head for multi-class prediction.
Unlike traditional ResNets in computer vision, our model omits convolutions.
We empirically found that removing all BatchNorm layers in FC-ResNet leads to improved performance of \TheName{}.

We train \TheName{} using an Adam optimizer (learning rate 5e-5 and weight decay 1e-5. For all training and evaluation of \TheName{}, we linearly ramp up $\lambda$ to its maximum value~\cite{tarvainen2017mean}.
We perform up-sampling through data augmentation to each malware family in the labeled set to create a balanced $\mathcal{X}^\prime$.
Each evaluation task is performed five times with different labeled sets, and we report the average performance of each model.

\begin{algorithm}[tbp]
\caption{\TheName{} data augmentation pipeline}
\label{alg:algorithm}
\small
\begin{algorithmic}[1] 
\REQUIRE  Malware feature set $\mathcal{S}=concat(\mathcal{S}_i, \mathcal{S}_n)$, non-interpolatable feature embedding set $\mathcal{H}_n$,  feature invariance embedding sets $\mathcal{H}_i^{sim}$ and $\mathcal{H}_n^{sim}$, pre-trained encoder $\phi_i$
\FOR{$s\in S$}
\STATE Randomly select a similar malware $s^\prime$ from the set of the top-k similar malware for $s$
\STATE Linearly combine the interpolatable features of $s$ and $s^\prime$ using Eq. (\ref{eq:mix-I}). Let $\Tilde{s}_i$ denote the result
\STATE Retrieve from $\mathcal{H}_n$ the projections of the non-interpolatable features of $s$ and $s'$, denote these projections $h_n$ and $h_n^\prime$ 
\STATE Linearly combine $h_n$ and $h_n^\prime$ using Eq. (\ref{eq:mix-U}) to obtain $\Tilde{h}_n$
\STATE Identify $k ~candidates$ in $\mathcal{H}_n$ most similar to $\Tilde{h}_n$
\FOR{$\hat{h}_n $ in $candidates$}
\STATE With $\hat{h}_n = \hat{h}_n^{sim}\oplus \hat{h}_n^{dis}$, set $\hat{h}_n = \hat{h}_n^{sim}$
\ENDFOR
\STATE Use $\phi_i$ to project $\Tilde{s}_i$ to $\mathcal{H}_i$. Let $\Tilde{h}_i$ denote the result
\STATE With $\Tilde{h}_i = \Tilde{h}_i^{sim}\oplus \Tilde{h}_i^{dis}$, set $\Tilde{h}_i = \Tilde{h}_i^{sim}$
\STATE Select $\hat{h}_n $ from $candidates$ s.t. $\hat{h}_n $ is most similar to $\Tilde{h}_i$
\STATE Identify the $\Tilde{s}_n\in \mathcal{S}_n$ that maps to $\hat{h}_n$. Use $\Tilde{s}_n$ as the mixture of $s$ and $s^\prime$ by their non-interpolatable features. 
\RETURN $\mathcal{A}(s) := \Tilde{s}_i\oplus \Tilde{s}_n$ 
\ENDFOR
\end{algorithmic}
\end{algorithm}

\begin{table}[b]
    \centering
    \setlength{\tabcolsep}{1mm}{
    \caption{Saturation evaluation of \TheName{} against MOTIF-50, with Fully Connected ResNet and LightGBM provided as a supervised standard for reference.
    \label{tab:saturation-motif}}
    \scalebox{0.86}{\begin{tabular}{l r c c c c}
        \toprule
        Models & Labels & Accuracy & Precision  & Recall & F1 \\
        \midrule
        \multirow{6}{*}{\TheName{}} 
         & $1\%$ & 40.45$\pm$4.77 & 45.45$\pm$6.59 & 43.06$\pm$5.31 & 39.70$\pm$5.16 \\
         & $5\%$ & 46.13$\pm$5.04 & 47.20$\pm$6.73 & 45.26$\pm$6.02 & 42.31$\pm$5.91 \\
         & $10\%$ & 49.33$\pm$5.46 & 51.52$\pm$3.42 & 46.71$\pm$3.26 & 44.71$\pm$4.17\\
         & $25\%$ & 63.96$\pm$2.78 & 64.22$\pm$2.77 & 62.44$\pm$2.664 & 59.82$\pm$3.00 \\
         & $50\%$ & 71.31$\pm$2.50 & 71.18$\pm$1.77 & 70.56$\pm$1.96 & 68.31$\pm$2.04 \\
         & $75\%$ & 74.18$\pm$1.23 & 75.81$\pm$2.10 & 74.08$\pm$1.48 & 71.59$\pm$1.28 \\
        \midrule
        FC-ResNet~\cite{he2016deep} & $100\%$ & 81.15 & 81.78 & 79.30 & 77.81 \\ 
        \midrule[0.01pt]
        LightGBM~\cite{ke2017lightgbm} & $100\%$ & 83.07 & 79.94 & 79.72 & 77.40 \\ 
        \bottomrule
    \end{tabular}}
    }
\end{table}

\section{Saturation Analysis on of MOTIF-50} \label{sec:appendix:saturation}

We performed an additional saturation analysis on MOTIF-50, analogous to that on BODMAS-20 in Section~\ref{sub:rq1}. We select $1\%$, $5\%$, $10\%$, $50\%$, and $75\%$ of all data in training set as labeled data, ensuring that at least one sample from each family is labeled. 
In the $1\%$ setting, almost all families contain only one-shot labels.
Table~\ref{tab:saturation-motif} shows that achieving satisfactory performance on MOTIF-50 requires a higher proportion of labeled data compared to BODMAS-20 (Table~\ref{tab:rq1}). 
However, we note that even with 25\% labeled data, most families in MOTIF-50 still remain in the few-shot regime (i.e. fewer than 5 labels per family) due to the larger number of families and the small family sizes in this dataset. 

More importantly, we hypothesize that \TheName{}'s degrading performance with fewer labels on MOTIF-50 may be more reflective of label quality issues rather than the true impact of the reduced label quantity. Specifically, MOTIF's label noise is reported to be as high as 40\%~\cite{wu2023grim}. As the number of labeled samples decreases, the model become more reliant on the quality of each label; in contrast, with larger labeled sets, the distribution of labeled samples within each family naturally converges to the true mean, even in the presence of high label noises. Thus, in few-shot settings, it is typically assumed that labels are highly accurate, as experts have time to carefully curate them. These factors suggest that MOTIF-50 may not be the ideal dataset for evaluating the impact of decreasing labeled data quantities on model performance.

Nonetheless, even under such settings, \TheName{} still achieves satisfactory performance under few-shot scenarios and outperforms baselines in near one-shot settings under high noise, as shown Section~\ref{sec:comparison}.


\section{Expressiveness of FC-ResNet} \label{sec:appendix:resnet}
Fully Connected ResNet (FC-ResNet), our chosen  base classifier, achieved performance comparable to LightGBM in fully supervised settings on both BODMAS-20 and MOTIF-50, as shown in Table~\ref{tab:rq1} and Table~\ref{tab:saturation-motif}. In tabular machine learning, gradient boosting techniques like LightGBM are often considered state-of-the-art due to their strong performance. In fact, LightGBM was the primary classification method in the original BODMAS paper~\cite{yang2021bodmas}. However, our semi-supervised framework requires a base classifier that outputs label probabilities, making GBDT (e.g. LightGBM) unsuitable for this role.

To explore alternative expressive models, we experimented with a Transformer-based classifier designed specifically for tabular malware detection. This model projects input features to a 128-dimensional latent space, applies positional encoding, and utilizes two Transformer encoder blocks with four attention heads and a hidden dimension of 256 in the feedforward layers. However, despite its theoretical expressiveness, the Transformer achieved only 91.16 Accuracy, 91.89 Precision, 92.80 Recall, and 92.22 F1 under fully supervised settings on BODMAS-20, underperforming compared to FC-ResNet (Table~\ref{tab:rq1}). This is possibly due to limtied dataset sizes and suggests that a more expressive neural model does not necessarily translate to better performance for this tabular malware classification task. 

As further evidence, prior studies have demonstrated the effectiveness of a simple three-layer MLP as a base classifier on similar tasks~\cite{wu2023grim}. Collectively, these findings confirm that FC-ResNet offers sufficient expressiveness to serve as the base classifier in our semi-supervised framework.

\section{Comparison Against kNearestNeighbor}\label{sec:appendix:knn}
A standard kNeighborsClassifier achieves 
$68.56\pm1.07$ accuracy and 
$65.63\pm1.10$ f1 (macro) on BODMAS-20 with 
$1\%$ labeled data, over MalMixer’s performance at 
$83.35\pm1.03$ and $84.60\pm1.31$. We note that, intuitively, malware classification relies on drawing boundaries in feature space. A simple similarity-based classifier does not know where to draw boundaries between sparsely distributed few-shot data. In contrast, \TheName{} makes use of the unlabeled data neighboring labeled data to narrow down margins around decision boundaries.

\section{Experiment on All BODMAS}\label{sec:appendix:all}

We performed two evaluations of \TheName{} on the entire BODMAS dataset (with 581 families). 
\squishlist
    \item \textbf{Experiment 1 (All Families Labeled)}. The BODMAS dataset contains many families with only one sample. After a random 80/20 train/test split, only 296 families are present in both train and test sets. To provide consistent evaluation metrics, we trained and tested on these 296 families only. We labeled $1\%$ of the samples, ensuring at least one labeled sample per family \TheName{} achieves $64.31$, $49.0$2, $54.28$, and $45.70$ in accuracy, precision, recall, and f1 (macro). For reference, the fully supervised FC-ResNet yielded performance metrics $86.93$, $63.29$, $61.05$, $60.36$. While \TheName{}’s performance drops substantially in the presence of numerous small families, this outcome is unsurprising given the challenge of handling more small families, especially when benchmarked against the also deteriorated performance of a fully supervised approach.
    \item \textbf{Experiment 2 (All Families, Only Top-100 Families Labeled)}. Here, we conducted a 80/20 random train/test split among all 581 malware families, so both sets contained families from outside the top-100 in prevalence. \TheName{} was trained using this data, with only $1\%$ of the top-100 prevalent families labeled. During testing, any family not within the top-100 was considered a False Positive if it was classified among those top-100 families. This set-up yielded performance metrics of $69.58$, $60.93$, $60.92$, and $56.11$ on the 100 labeled families. For context, in a prior experiment where both train and test sets solely contained the top-100 families, the performance metrics stood at $70.65$, $63.60$, $62.11$, and $58.78$. These results suggest that \TheName{}'s performance on large families remains strong even with False Positives from small, unidentified families. In general, the presence of small families does not significantly impact the model's performance on large families. We hypothesize that the minimal size of these long-tail families limits their overall impact on both training and inference metrics.

\squishend

We note that accurately classifying small families is not the intended application of MalMixer. In practice, when faced with a surge of new, unlabeled malware data, practitioners can promptly label only a limited but representative subset, expediting the analysis of large families. Such preliminary classifications into larger families can still provide insightful information about the nature of the unlabeled samples and free up engineers’ time to analyze the smaller families.


\section{Obfuscation Experiment}\label{sec:appendix:obfuscation}

To evaluate \TheName{}'s performance under code obfuscation, we used the UCSB-Packed dataset~\cite{aghakhani2020malware}, which employs 10 unique packers to create obfuscated malware samples. As the dataset does not have malware family labels, we turned to VirusTotal reports to assign family labels to each sample. We limited our dataset to only the top 20 most common families, and perform a random 80/20 train-test split. We compare \TheName{} against baseline models using only $1\%$ labeled sample in the train set. 

Under packing, many of the $k$-NN and clustering assumptions outlined in Sections~\ref{sec:intro} and \ref{sec:results-rq0} become weaker, negatively impacting \TheName{}'s performance on the UCSB-Packed dataset. However, as shown in Table~\ref{tab:obfuscation}, \TheName{} remains among the top two methods, performing slightly worse in accuracy and precision but better in recall and F1 compared to MixMatch (Gaussian). 
Notably, MixMatch (Gaussian) is structurally identical to \TheName{} but replaces our proposed data augmentation with Gaussian noise augmentation.
This suggests that even when some of \TheName{}'s underlying assumptions are violated, our proposed data augmentation strategy remains at least competitive with Gaussian-noise-based augmentation, which does not respect malware feature semantics.

It’s vital to recognize that the performances in Table~\ref{tab:obfuscation} only partially capture the increased challenges in classifying obfuscated samples. The noises, inconsistencies, and inaccuracies inherent in VirusTotal’s family labels further complicates matters~\cite{yang2021bodmas}. Experimentally, we’ve noted that even a fully supervised FC-ResNet achieved only 72.46 accuracy on this dataset. This calls for future research efforts on curating obfuscated datasets with less noisy family labels, as well as on improving static-feature-based, few-shot, and semi-supervised classifiers' robustness to obfuscation.


\begin{table}[tbp] 
    \centering
    \setlength{\tabcolsep}{1mm}{
    \caption{Comparison of \TheName{} against semi-supervised methods on obfuscated samples with $1\%$ labeled data. 
    \label{tab:obfuscation}}
    \scalebox{0.86}{\begin{tabular}{l c c c c} 
        \toprule
        Models & Accuracy & Precision & Recall & F1 \\ 
        \midrule
        VAT & 40.68$\pm$0.81 & 32.03$\pm$3.67 & 23.06$\pm$1.61 & 22.47$\pm$2.05 \\
        VAT (aug) & 38.89$\pm$0.86 & 32.13$\pm$1.13 & 31.93$\pm$1.61 & 29.60$\pm$0.35 \\
        MixUp & 42.55$\pm$1.02 & 35.53$\pm$2.14  & 29.33$\pm$1.20 & 28.89$\pm$0.86 \\
        MixMatch & 43.31$\pm$1.10 & 37.03$\pm$1.02 & 31.28$\pm$1.37 & 31.35$\pm$1.19 \\
        MixMatch (Gaussian) & 46.80$\pm$0.73 & 39.43$\pm$2.16 & 34.84$\pm$0.97 & 34.07$\pm$0.25 \\
        \midrule[0.01pt]
        MORSE & 43.59$\pm$2.31 & 42.88$\pm$2.84 & 27.77$\pm$2.38 & 28.62$\pm$2.89 \\
        \midrule
        \TheName{} & 46.08$\pm$0.76 & 37.41$\pm$0.74 & 35.70$\pm$0.69 & 35.24$\pm$0.53 \\
        \bottomrule
    \end{tabular}}
    }
\end{table}

\section{Interpolatable vs. Non-interpolatable feature sets for EMBER features}\label{sec:appendix:interpolatable}
\noindent \textbf{Interpolatable Features}: \{ByteHistogram, ByteEntropyHistogram~\cite{saxe2015deep}, frequency of printable characters, occurrences of strings (e.g. ``HTTP:/'', ``MZ''), length of the bytes of the file, virtual size of the LIEF-parsed binary, number of exported/imported functions, number of symbols, size of code/headers/heap commits, total number of sections, number of sections with a certain feature (e.g. zero size, empty name), size/virtual size of data directories\}

\noindent \textbf{Non-interpolatable Features}: \{whether the binary has a certain feature (e.g. debug section, Relocation, is signed), hash COFF machine type/characteristics, hash optional subsystem/dll characteristics/magic number, major/minor image/linker/operating system/subsystem version, hash on pair of section name and size/entropy/virtual size, hash on name of entry point, hash on list of properties of entry point, hash on list of unique imported libraries, hash on list of library:function\}

\section{Train-Test Overlap of Non-Interpolatable Features in Experimental Datasets}\label{sec:appendix:noninterpolatable-overlap}

One important detail about our experimental dataset is the train-test overlap of non-interpolatable features. This information could clarify whether the models are relying too much on non-interpolatable features..

For BODMAS-20, 81\% of non-interpolatable features within the same malware family overlap between the training and test sets. In contrast, the same number for MOTIF-50 is just 6\%. This partially explains why all models, including \TheName{} and baselines, performed better on BODMAS-20. However, \TheName{} outperforms other models on both datasets, showing its ability to handle both high and low train-test overlaps in non-interpolatable features. This reasonably suggests that our model does not overly rely on non-interpolatable features, at least compared to the other baseline models.

Furthermore, 37\% of non-interpolatable features are unique per malware family in BODMAS-20. The same number for MOTIF-50 is 97\%. These numbers suggest that our data augmentation pipeline for non-interpolatable features can indeed replace one sample’s non-interpolatable features with other distinct non-interpolatable feature sets, thus introducing diversity to mitigate overfitting and increase robustness.

\begin{algorithm}[tbp]
\caption{The modified MORSE algorithm.}
\label{alg:morse}
\small
\begin{algorithmic}[1]
\REQUIRE Training dataset $D$, number of total training epochs $K$, \st{labeled data's proportion $d$, starting re-weighting epoch $T_d$,} confidence threshold $\tau$, unsupervised loss weight $\lambda$, learning rate $\alpha$.
\STATE \st{\textbf{Initialization:} Initialize the weights $\Theta$ for the model $f(\cdot)$ by using the entire dataset $D$ with only a few epochs.}
\FOR{$k = 0, 1, 2, \dots, K$}
    \STATE \st{Partition the training dataset $D$ into labeled dataset $\mathcal{X}$ and unlabeled dataset $\mathcal{U}$: select the top $d\%$ examples with the least loss values from each given class and treat them as labeled data, and the rest as unlabeled data.}\\
    \hl{Use the groundtruth labeled dataset $\mathcal{X}$ and unlabeled dataset $\mathcal{U}$.}
    \FOR{$iter = 1, 2, \dots, num\_batches$}
        \STATE From $\mathcal{X}$, draw a mini-batch:\\ $\{(\mathbf{x}_b, y_b) : b \in (1, \dots, B)\}$
        \STATE From $\mathcal{U}$, draw a mini-batch:\\ $\{\mathbf{u}_b : b \in (1, \dots, B_\mu)\}$
        \FOR{$b = 1, 2, \dots, B_\mu$}
            \STATE // Perform weak augmentation against $\mathbf{u}_b$
            \STATE $\mathbf{q}_b = f(g(\mathbf{u}_b); \Theta)$
            \STATE $\widehat{q}_b = \arg\max(\mathbf{q}_b)$
        \ENDFOR
        \IF{\st{$k < T_d$} \hl{True}}
            \STATE // Perform weak augmentation against $\mathbf{x}_b$
            \STATE $\mathcal{L}_s = \frac{1}{B} \sum_{b=1}^{B} H(y_b, f(g(\mathbf{x}_b)))$
            \STATE // Perform strong augmentation against $\mathbf{u}_b$
            \STATE $\mathcal{L}_u = \frac{1}{B_\mu} \sum_{b=1}^{B_\mu} \textbf{I}(\max(\mathbf{q}_b) > \tau) H(\widehat{q}_b, f(h(\mathbf{u}_b)))$
        \ELSE
            \STATE \st{// Calculate the weight of each training sample}
            \STATE \st{// Perform weak augmentation against $\mathbf{x}_b$}
            \STATE \st{$\mathcal{L}_s = \frac{1}{B} \sum_{b=1}^{B} w_b H(y_b, f(g(\mathbf{x}_b)))$}
            \STATE \st{// Perform strong augmentation against $\mathbf{u}_b$}
            \STATE \st{$\mathcal{L}_{u1} = \frac{1}{B_\mu} \sum_{b=1}^{B_\mu} \textbf{I}(\max(\mathbf{q}_b) > \tau)$}
            \STATE \st{$\mathcal{L}_u = \mathcal{L}_{u1} + w_b H(\widehat{q}_b, f(h(\mathbf{u}_b)))$}
        \ENDIF
        \STATE $\mathcal{L} = \mathcal{L}_s + \lambda \mathcal{L}_u$
        \STATE Update the model’s weights $\Theta$ by minimizing the loss function $\mathcal{L}$
        \STATE Optional: decay the learning rate $\alpha$
    \ENDFOR
\ENDFOR
\STATE \textbf{Output:} the well-trained model $f(\cdot; \Theta)$.
\end{algorithmic}
\end{algorithm}

\section{Adaptations of MORSE}\label{sec:appendix:morse}
MORSE (\textbf{M}alware c\textbf{O}rpus fo\textbf{R} \textbf{S}emi-supervised l\textbf{E}arning)~\cite{wu2023grim} is a state-of-the-art malware family classification algorithm, originally designed to address challenges with incorrect groundtruth malware labels and class imbalance. Leveraging semi-supervised learning, MORSE gracefully handles malware classifications in label-scarce scenarios using both labeled and unlabeled data. 

At a high level, MORSE's original design starts with a fully labeled dataset but retains only the most confidently classified samples as labeled data, treating the rest as unlabeled. Then, the labeled and unlabeled data is fed into the FixMatch~\cite{sohn2020fixmatch} pipeline, a semi-supervised learning approach originally developed for image classification and a successor to MixMatch~\cite{berthelot2019mixmatch}. Within this pipeline, MORSE introduces sample re-weighting~\cite{cui2019class} to mitigate class imbalance. Additionally, it employs a malware-specific data augmentation technique: an augmented sample is obtained by swapping a subset of randomly selected features from a given sample with those from another randomly selected sample. To align with FixMatch’s requirement for both weak and strong augmentations, MORSE defines weak augmentation as replacing $10\%$ of features and strong augmentation as replacing $20\%$. Note that although MORSE proposes a malware-specific data augmentation pipeline, the random swapping still largely disregards the underlying semantics of each feature.

Unlike \TheName{}, MORSE utilizes a 3-layer MLP as its base classifier. We chose not to replace it with our FC-ResNet, as our experiments showed no performance gains from this substitution.

To adapt MORSE to our experimental settings, we note that our few-shot settings, with already scarce labels, do not permit the model to decide what data should be considered labeled vs. unlabeled. Furthermore, the few-shot settings have rendered the re-sampling techniques empirically unhelpful, partially due to difficulties with inferring class sizes based on few-shot labels. As a result, we have removed these features from MORSE, and have detailed our changes to the original MORSE framework in Algorithm~\ref{alg:morse}, where steps removed are marked with strikethrough and added steps highlighted in gray.

With careful fine-tuning, we found that strictly following the hyperparameter set used in MORSE's original experiments --- which were conducted on their synthetic dataset curated from a subset of BODMAS --- while only adjusting the confidence threshold $\tau$ to $0.95$, yielded the best results in our experimental setting.

\begin{table*}[tbp]
    \centering
    \setlength{\tabcolsep}{2.5mm}{
    \caption{Results grouped by family from the BODMAS-20 experimental design with $1\%$ of labeled data. We note that the few families with lower f1 scores are intrinsically harder to classify, even by the standards of a fully-supervised model.
    \label{tab:group}}
    \scalebox{1}{\begin{tabular}{l l r l l l}
        \toprule
Malware Family & Description  & Proportion of Dataset & Precision & Recall & F1 \\
\midrule
AutoIt & Scripting-language-based & 2.44\% & $92.38\pm2.63$ & $95.21\pm5.74$ & $89.76\pm3.20$ \\
Benjamin & Worm & 2.69\% & $100.00\pm0.00$ & $100.00\pm0.00$ & $100.00\pm0.00$ \\
Berbew & Backdoor/Trojan & 4.40\% & $100.00\pm0.00$ & $95.62\pm6.84$ & $83.08\pm10.10$ \\
CeeInject & Virus/Trojan & 2.94\% & $100.00\pm0.00$ & $84.53\pm7.11$ & $100.00\pm0.00$ \\
Dinwod & Trojan & 5.17\% & $74.95\pm13.25$ & $83.42\pm10.55$ & $62.68\pm6.93$ \\
Drolnux & Worm/Trojan & 2.31\% & $100.00\pm0.00$ & $99.79\pm0.26$ & $100.00\pm0.00$ \\
GandCrab & Ransomware & 2.41\% & $97.81\pm3.00$ & $92.80\pm2.03$ & $97.81\pm3.00$ \\
Ganelp & Worm & 5.61\% & $84.24\pm13.84$ & $96.64\pm3.12$ & $75.52\pm11.81$ \\
Gepys & Trojan & 2.83\% & $99.10\pm1.81$ & $85.91\pm13.03$ & $97.38\pm5.24$ \\
Mira & Trojan/Ransomware & 4.93\% & $100.00\pm0.00$ & $97.48\pm0.72$ & $99.60\pm0.43$ \\
Musecador & Trojan & 2.65\% & $100.00\pm0.00$ & $98.85\pm1.47$ & $99.91\pm0.18$ \\
Quakart & Spyware & 2.07\% & $55.88\pm9.37$ & $100.00\pm0.00$ & $49.84\pm9.74$ \\
Sfone & Worm & 11.89\% & $100.00\pm0.00$ & $98.91\pm0.00$ & $100.00\pm0.00$ \\
Sillyp2p & Worm/Trojan & 4.06\% & $96.47\pm4.29$ & $99.27\pm0.36$ & $91.52\pm9.19$ \\
Small & Downloader/Trojan & 8.40\% & $98.76\pm0.73$ & $97.56\pm1.69$ & $98.07\pm0.99$ \\
SmokeLoader & Backdoor/Trojan & 2.18\% & $94.53\pm4.94$ & $96.59\pm1.14$ & $91.16\pm5.21$ \\
Unruy & Trojan & 2.18\% & $100.00\pm0.00$ & $100.00\pm0.00$ & $100.00\pm0.00$ \\
Upatre & Backdoor/Trojan & 9.81\% & $99.44\pm4.59$ & $98.55\pm1.39$ & $90.32\pm5.36$ \\
Wabot & Backdoor & 9.24\% & $98.81\pm1.46$ & $99.81\pm0.07$ & $98.14\pm2.19$ \\
Wacatac & Ransomware & 11.80\% & $73.05\pm12.21$ & $73.20\pm17.86$ & $57.00\pm7.22$ \\
\bottomrule
\end{tabular}}
    }
\end{table*}

\end{appendices}

\end{document}